\documentclass[twocolumn]{aastex7}

\usepackage{amsmath,amsthm}
\usepackage{booktabs}
\usepackage{threeparttable}
\usepackage{tabularx}
\usepackage{xcolor}
\usepackage{times}
\usepackage{graphicx}
\usepackage{comment}

\newcommand{\Lx}{L_{\rm X}}
\newcommand{\Lxf}{L_{\rm X,5}}
\newcommand{\Lk}{L_{\rm K}}
\newcommand{\Tx}{T_{\rm  X}}
\newcommand{\avTx}{\langle\Tx\rangle}
\newcommand{\avTf}{\langle T_{\rm X,5} \rangle}
\newcommand{\SigX}{\Sigma_{\rm X}}
\newcommand{\Mst}{M_*}
\newcommand{\dMst}{\dot M_*}
\newcommand{\rhost}{\rho_*}
\newcommand{\rst}{r_*}
\newcommand{\qs}{q_*}
\newcommand{\Reff}{R_{\rm e}}
\newcommand{\MR}{{\cal R}}
\newcommand{\Mgal}{M_{\rm g}}
\newcommand{\rgal}{r_{\rm g}}
\newcommand{\rhogal}{\rho_{\rm g}}
\newcommand{\Mbh}{M_{\rm BH}}
\newcommand{\vh}{v_{\rm h}}
\newcommand{\rh}{r_{\rm h}}
\newcommand{\rhoh}{\rho_{\rm h}}
\newcommand{\csih}{\xi_{\rm h}}
\newcommand{\Msol}{{\rm M}_\odot}
\newcommand{\kpc}{{\rm kpc}}
\newcommand{\kms}{{\rm km}\;{\rm s}^{-1}}
\newcommand{\MDM}{M_{\rm DM}}
\newcommand{\Dels}{\Delta_*}
\newcommand{\sigmas}{\sigma_*}
\newcommand{\phiBH}{\phi_{\rm BH}}
\newcommand{\vphib}{\overline{v_\varphi}}
\newcommand{\sigmaphi}{\sigma_\varphi}
\newcommand{\ke}{k_{\rm e}}
\newcommand{\taunu}{\tau_{\nu}}
\newcommand{\taunup}{\tau_\nu^\prime}
\newcommand{\spr}{s^\prime}
\newcommand{\Inu}{I_{\nu}}
\newcommand{\alphanu}{\alpha_{\nu}}
\newcommand{\jnu}{j_{\nu}}
\newcommand{\Snu}{S_{\nu}}

\newcommand{\zmin}{z_{\rm min}}
\newcommand{\zmax}{z_{\rm max}}
\newcommand{\Signu}{\Sigma_{\nu}}
\newcommand{\numin}{\nu_{\rm min}}
\newcommand{\numax}{\nu_{\rm max}}
\newcommand{\QT}{Q_{\rm T}}
\newcommand{\cD}{c_{\rm D}}
\newcommand{\Vrot}{V_{\rm rot}}
\newcommand{\SigmaT}{\Sigma_{\rm T}}
\newcommand{\kr}{\kappa_R}
\newcommand{\rJ}{r_{\rm J}}
\newcommand{\MJ}{M_{\rm J}}
\newcommand{\rhoJ}{\rho_{\rm J}}
\newcommand{\rhoi}{\rho_i}
\newcommand{\Ti}{T_i}
\newcommand{\Vi}{V_i}
\newcommand{\rhoit}{\tilde\rhoi}

\newcommand{\Vit}{\tilde\Vi}
\newcommand{\avL}{L_{\rm multi}}
\newcommand{\avT}{T_{\rm multi}}
\newcommand{\rhohi}{\rho_{\rm high}}
\newcommand{\rholo}{\rho_{\rm low}}
\newcommand{\Vh}{V_{\rm high}}
\newcommand{\Vl}{V_{\rm low}}

\begin{document}

\shortauthors{S. Pellegrini, L. Ciotti, Z. Gan, D.-W. Kim,  J.P. Ostriker} 
\shorttitle{X-ray halos, AGN feedback and CGM accretion in ETGs}

\title{X-ray Halos of Early-Type Galaxies with AGN Feedback and Accretion from a Circumgalactic Medium: models and observations}

\author[orcid=0000-0002-8974-2996,sname=Pellegrini]{Silvia  Pellegrini}
\affiliation{Department of Physics and Astronomy, University of Bologna, via Gobetti 93/2, I-40129 Bologna, Italy}
\affiliation{INAF-Osservatorio di Astrofisica e Scienza dello Spazio di Bologna,   Via Gobetti 93/3, I-40129 Bologna, Italy}
\email[show]{silvia.pellegrini@unibo.it}

\author{Luca Ciotti}
\affiliation{Department of Physics and Astronomy, University of  Bologna, via Gobetti 93/2, I-40129 Bologna, Italy}
\email{luca.ciotti@unibo.it}

\author{Zhaoming Gan}
\affiliation{New Mexico Consortium, Los Alamos, NM 87544, USA} 
\email{ganzhaoming@gmail.com}

\author{Dong-Woo Kim}
\affiliation{Harvard-Smithsonian Center for Astrophysics, 60 Garden Street, Cambridge, MA 02138, USA}
\email{dkim@cfa.harvard.edu}

\author {Jeremiah P. Ostriker}
\affiliation{Department of Astronomy, Columbia University, 550 West 120th St, New York, NY 10027, USA}
\affiliation{ Department of Astrophysical Sciences, Princeton University, Princeton, NJ 08544, USA}
\email{ostriker@princeton.edu}

\correspondingauthor{Silvia Pellegrini}

\begin{abstract}
The knowledge of the X-ray properties of the hot gas halos of early-type galaxies has significantly advanced in the past years,
for large and homogeneously investigated samples.
We compare these results with the X-ray properties of an exploratory set of gas evolution models in realistic early-type galaxies,
produced with our high resolution 2D hydrodynamical code MACER that includes AGN feedback and accretion from a circumgalactic medium.
The model X-ray emission and absorption are integrated along the line of sight, to obtain maps of
the surface brightness $\SigX$ and temperature $\Tx$. The X-ray diagnostics considered are the luminosity
and average temperature for the whole galaxy ($\Lx$ and $\avTx$) and within 5 optical effective radii ($\Lxf$ and $\avTf$), and 
the circularized profiles $\SigX (R)$ and $\Tx(R)$. The values for $\Lx$, $\Lxf$, $\avTx$, and $\avTf$ compare very well with those observed.
The $\SigX (R)$ and $\Tx(R)$ also present qualitative similarities with those of the representative galaxy NGC5129, and of ETGs
  with the most commonly observed shape for $\Tx(R)$: 
$\SigX(R)$ matches the observed profile over many optical effective radii $\Reff$, and $\Tx(R)$ reproduces the 
characteristic bump that peaks at $R=(1\div 3)\Reff$. Inside the peak position,  $\Tx(R)$ declines towards the center, but the explored
models are systematically hotter by $\simeq 30\%$; possible explanations for this discrepancy are discussed.
Interestingly,  $\SigX (R)$ and $\Tx(R)$ as large as observed outside of $R\simeq \Reff$  are reproduced only with
significant accretion from a circumgalactic medium, highlighting its importance.
\end{abstract}


\section{Introduction}\label{Intro}
In the past years, the X-ray properties of the hot ISM of Early-Type Galaxies (hereafter ETGs) have been deeply investigated, thanks 
to the data obtained with the $Chandra$ and XMM-$Newton$ observatories (Boroson et al. 2011, Kim \& Fabbiano 2015, Goulding et al. 2016, Lakhchaura et al. 2018,
Babyk et al. 2018, Islam et al. 2021; Nardini et al. 2022, and references therein). In particular, the data of 70 ETGs in the $Chandra$ archive were homogeneously and
extensively analyzed, and the resulting hot gas properties collected in the $Chandra$ Galaxy Atlas (Kim et al. 2019, hereafter K19). Among many products,  for each
galaxy this Atlas provides the X-ray luminosity $\Lx$ and the average temperature $\avTx$ for the whole galactic extent and within representative radii (e.g., $\Reff$,
the optical effective radius, and 5$\Reff$), and the X-ray surface brightness and temperature profiles $\SigX (R)$ and $\Tx(R)$. One major outcome was the recognition
that most temperature profiles fit in a ``universal'' shape (Kim et al. 2020, hereafter K20). Except for a set of ETGs  (13\% of the sample) with $\Tx(R)$ monotonically
declining outwards, for most ETGs  (82\%)  the $\Tx(R)$ profile fits in a description that includes a broad bump at intermediate radii, with a maximum $\Tx$ located
at $(1\div 3)\Reff$, and a decline both inward and outward.
Inside a few kpc, $\Tx (R)$  can either keep declining down to the innermost observed point, or become flat, or show a central increase. The features in
this universal profile have been attributed quite naturally to the roles of the environment, for the outer galactic regions, and of the AGN feedback for the central ones.

On the modeling side,  progress has been stimulated by the observational results, and by advances in the
hydrodynamical simulations (e.g., Choi et al. 2015, Gaspari et al. 2017, Ciotti et al. 2017, Pellegrini et al. 2018, Wang et al. 2019, Gan et al. 2019a, Li et al. 2020, 
Truong et al. 2020, Mohapatra et al. 2025).
However, no study so far performed a close comparison between the results obtained for the hot gas by simulations especially designed for realistic ETGs,  covering a range
of their main properties, and what observed for the gas in the X-rays, and collected in the large studies mentioned above.
We investigate here to what extent the gas properties of the archival $Chandra$ ETGs are reproduced by the modeling of the
gas evolution with our high resolution 2D hydrodynamical code MACER.
This modeling includes mass and energy sources from an old stellar population, mechanical and radiative heating from a central AGN, and
also important phenomena such as galaxy rotation, star formation, and cosmological inflow from a circumgalactic environment (CGM).
The code and the input physics have been developed by Ciotti \& Ostriker (2001, 2007, 2012) and collaborators, with recent major upgrades by Gan et al. (2019a, hereafter G19a; Gan et al. 2019b, 2020).
In these simulations the inner boundaries range from 2.5 pc to 25 pc to resolve the Bondi radius; while only
performed in 2D, they greatly exceed the spatial resolution available in most cosmological simulations.
For a comparison with X-ray observations, we use the set of simulations presented in Ciotti et al. (2022, hereafter C22).
C22 built realistic dynamical models for the host galaxies, including the possible presence of a group dark matter halo,
for a range of stellar masses and internal stellar kinematics. They used the latest version of the MACER code, improved in particular on the physical
treatment of AGN feedback, and of star formation and disk instabilities; also considered was the time evolution of the gravitational field of the stellar disk produced
by the rotating cooled gas, of the growing central supermassive black hole (hereafter SMBH), and of the stellar part of the galaxy due to the mass loss of stars; the
SMBH growth and the stellar mass loss also determine a time evolution of the stellar velocity dispersion and rotational velocity fields.
The effects of the presence of dust, and of a variable metal abundance, whose evolution is separatley followed for a number of metal species, were also included (Gan et al. 2020, Pellegrini et al. 2020). 
A preliminary overview of the hot gas properties of these simulations was given in C22; we focus here on a close comparison of these properties with
those observed by $Chandra$. For the models, we estimate global quantities as 
the X-ray luminosity and the average temperature computed over the whole galaxy ($\Lx$ and $\avTx$) and within 5$\Reff$ ($\Lxf$ and $\avTf$), and more detailed
properties as the surface brightness  profile $\SigX (R)$ and the temperature profile $\Tx(R)$. We find that the global model properties reproduce well those observed
for $Chandra$ ETGs.
The $\SigX (R)$ and $\Tx(R)$ profiles were first compared with those of a bright and well studied galaxy (NGC5129),
  that is the prototypical example of the most commonly observed temperature profile (43\% of the cases) in the classification of $\Tx(R)$ 
  made by K20 for 60 ETGs.  In this class of profiles, called ``hybrid-bump'' (hereafter HB), 
  the temperature inside the broad peak keeps declining down to the innermost observed radius.
For a few models of the most massive family, that turned out to be structurally similar to NGC5129, the $\SigX (R)$ shape compares well
with that of NGC5129, and their $\Tx(R)$ shows the characteristic observed bump; however, 
within a few kpc, the model temperature is larger than observed by $\simeq 30\%$.
This comparison further  indicated the strong importance of CGM accretion to reproduce the outer observed parts of NGC5129.  
Then we extended the analysis to the $\Tx(R)$ of all ETGs in the HB class and to the $\SigX (R)$ of those ETGs, among these,
  with optical luminosity similar to that of the successful models; the results obtained from the comparison with NGC5129 were confirmed, and their validity then
  extended. In particular,
the discrepancy in the behavior of the inner temperature appeared to be general; this highlighted the need for a
wider exploration of the parameters describing AGN accretion and feedback, and/or for the inclusion of 3D multiphase effects.

The paper is organized as follows: in Section 2 we briefly describe the models and the simulations; in Section 3, we compare $\Lx$, $\Lxf$
$\avTx$ and $\avTf$ of the models with those of large samples of ETGs observed with $Chandra$;
in Section 4 we investigate similarities and differences of the model $\SigX (R)$
and $\Tx(R)$ with those of the representative galaxy NGC5129 and of ETGs in the HB class;
in Section 5 we discuss the results and present the conclusions.

\section{The galaxy models}\label{mod}

A full description of the structure and dynamical properties of the axisymmetric 
ellipsoidal galaxy models used in the simulations, the input physics, and the numerical
implementation of all physical ingredients, is given in C22.
Below, we summarize some basic information.

\renewcommand\arraystretch{1.4}
\begin{table*}
\vspace{2mm}
\centering 
\caption{Structural properties of the models}
\vspace{2mm}
\begin{threeparttable}
\begin{tabularx}{\textwidth}{c@{\extracolsep{\fill}}cccccccccc}
\toprule 
  Model   &  $\Lk$       &  $\Mst$          & $\rst$   & $\Reff$    & $\sigmas(0)$&  $\vh$\\
  family   & $(10^{11}\,L_{K,\odot})$ &$(10^{11}\,\Msol)$& $(\kpc)$&$(\kpc)$  &  $(\kms)$      &  $(\kms)$   \\
                  & (1)         &      (2)              & (3)         &      (4)         &   (5)         &  (6)   \\ 
 \midrule                                                 
  LM  & $1.30$ & $1.54$ & $7.33$     & $4.57$  &  $223$  & $360$  \\
  \midrule 
  MM  & $2.65$ & $3.35$  & $11.29$  & $7.04$  &  $265$ & $427$ \\                                                                                                  
 \midrule 
  HM & $5.62$ & $7.80$ & $18.94$    & $11.80$  &  $312$& $504$ \\
  \bottomrule 
\end{tabularx}
\vspace{1.5mm}
\begin{tablenotes}[para,flushleft]
  \footnotesize For each model family, columns give: 
  (1) the galaxy luminosity in the K-band, 
  (2) the initial stellar mass, 
  (3) the scale-length of the stellar distribution (Equation \ref{eq:rhos}), 
  (4) the edge-on circularized effective radius, 
  (5) the central stellar velocity dispersion in absence of the SMBH, 
  (6) the asymptotic circular velocity of the quasi-isothermal DM 
  halo (Equation \ref{eq:rhoh}).  The models were built to lie on the Fundamental Plane of ETGs.
    \end{tablenotes}
    \end{threeparttable}
    \label{models}
    \vspace{5mm}
\end{table*}
\renewcommand\arraystretch{1.}

\subsection{The galaxy structure}

The stellar density distribution is an oblate ellipsoidal Jaffe (1983) model
of total mass $\Mst$, scale-length $\rst$, and minor-to-major axial ratio
$\qs$:
\begin{equation}
      \rhost(m)= 
      {\Mst\over 4\pi \qs\rst^3 m^2 (1+m)^2},
      \quad 
      m^2 \equiv {R^2\over\rst^2}+{z^2\over\qs^2\rst^2}.
\label{eq:rhos}
\end{equation}
In all the C22 simulations $\qs=0.7$ is adopted, corresponding to E3 galaxies when seen edge-on.
The effective radius of a model observed face-on is $\Reff^{\rm FO}\simeq 3\rst/4$,
and the circularized effective radius of the same model seen edge-on is
$\Reff=\sqrt{\qs}\,\Reff^{\rm FO}\simeq 0.63\rst$.
The stellar distribution is embedded in a galactic
dark matter (hereafter DM) halo; the stars plus DM galaxy density $\rhogal$ is a spherical Jaffe
distribution of total mass $\Mgal=\MR\Mst$ and scale length $\rgal=\xi\rst$, with $\xi \geq 1$:
\begin{equation}
      \rhogal(r)= 
      {\Mst\MR\xi\over 4\pi\rst^3 s^2 (\xi+s)^2},
      \quad 
      s \equiv {r\over\rst},
\label{eq:rhog}
\end{equation}
where $r=\sqrt{R^2+z^2}$ is the spherical radius.
C22 adopted $ \MR =\xi/\qs$ so that $\rho_{\rm DM}=\rhogal -\rhost$ reproduces the NFW profile over a large radial range
(Ciotti et al. 2021); in particular, $\MR\simeq 18$ and $\xi\simeq 12.6$ were taken.
In order to account for the effects of a group/cluster DM halo on the gas flows, the
models are also embedded in a spherically symmetric quasi-isothermal DM
halo, of asymptotic circular velocity $\vh$, scale-length $\rh=\csih\rst$, and density:
\begin{equation}
      \rhoh(r)=
      {\vh^2\over 4\pi G\rst^2(\csih^2 +s^2)}.
\label{eq:rhoh}
\end{equation}
We adopt $\csih =5$ and $\vh^2=2.6\sigmas^2$, where $\sigmas$ is the central stellar 
velocity dispersion due to $\rhogal$ only, so that $\rhoh$
is dynamically important only outside several $\Reff$. 
Finally, a SMBH of initial mass $\Mbh=\mu (0)\Mst = 10^{-3} \Mst$ is added at the center of the galaxy; this provides a time-evolving
potential $\phiBH(r,t)= -G\Mst\mu (t)/ r$, consequence of SMBH accretion.

C22 adopted three values for the initial stellar mass, i.e.,  $\Mst = 1.54\times 10^{11}\Msol$, 
$3.35\times 10^{11}\Msol$, and $7.80\times 10^{11}\Msol$, that correspond to the LM, MM, and HM families
of models. For all models the dark mass fraction 
$\MDM(r)/\Mgal(r)$ is $\simeq 0.52$ for $r=\Reff$, and $\simeq 0.64$ for $r=2\Reff$. 

The stellar velocity dispersion and the ordered velocity field $\vphib$ are determined as described in C22.
The equal vertical and radial components of the stellar velocity dispersion $\sigmas$, and the quantity  
$\Dels=\vphib^2+\sigmaphi^2 -\sigmas^2$ (where $\sigmaphi$ is the azimuthal dispersion) are obtained
from the Jeans equations. Then, $\vphib$ is given by a generalised Satoh (1980) $k$-decomposition:
\begin{equation}
      \vphib=k\,\sqrt{\Dels},
      \quad
      \sigmaphi^2=\sigmas^2+(1-k^2)\Dels.
\label{eq:vphis}
\end{equation}
For each galaxy mass $\Mst$, three types of rotation fields were implemented: non-rotating ($k=0$) galaxies,
where the stellar flattening $\qs$ is totally produced by $\sigmaphi$; fast-rotating, isotropic ($k=1$) galaxies, with the
flattening totally supported by ordered rotation; and galaxies with a spatially-dependent Satoh parameter
\begin{equation}
k_{\rm e}(r)={\rm e}^{-r/\Reff}.
\label{eq:ke}
\end{equation}
In this way, a total of nine galaxy models were studied. 
The main properties of the three families of models are summarized in Table~\ref{models}, and the nine models are listed in Table~\ref{tab2}.

\subsection{The input physics and the hydrodynamical simulations}

The input physics of the models (e.g., AGN feedback, star
formation, disk instabilities), and the numerical treatment of the
hydrodynamical equations are described in Section 3 of
C22. Here we recall their main characteristics.

\begin{figure}
\vskip -1.5truecm
\includegraphics[width=1\linewidth, keepaspectratio]{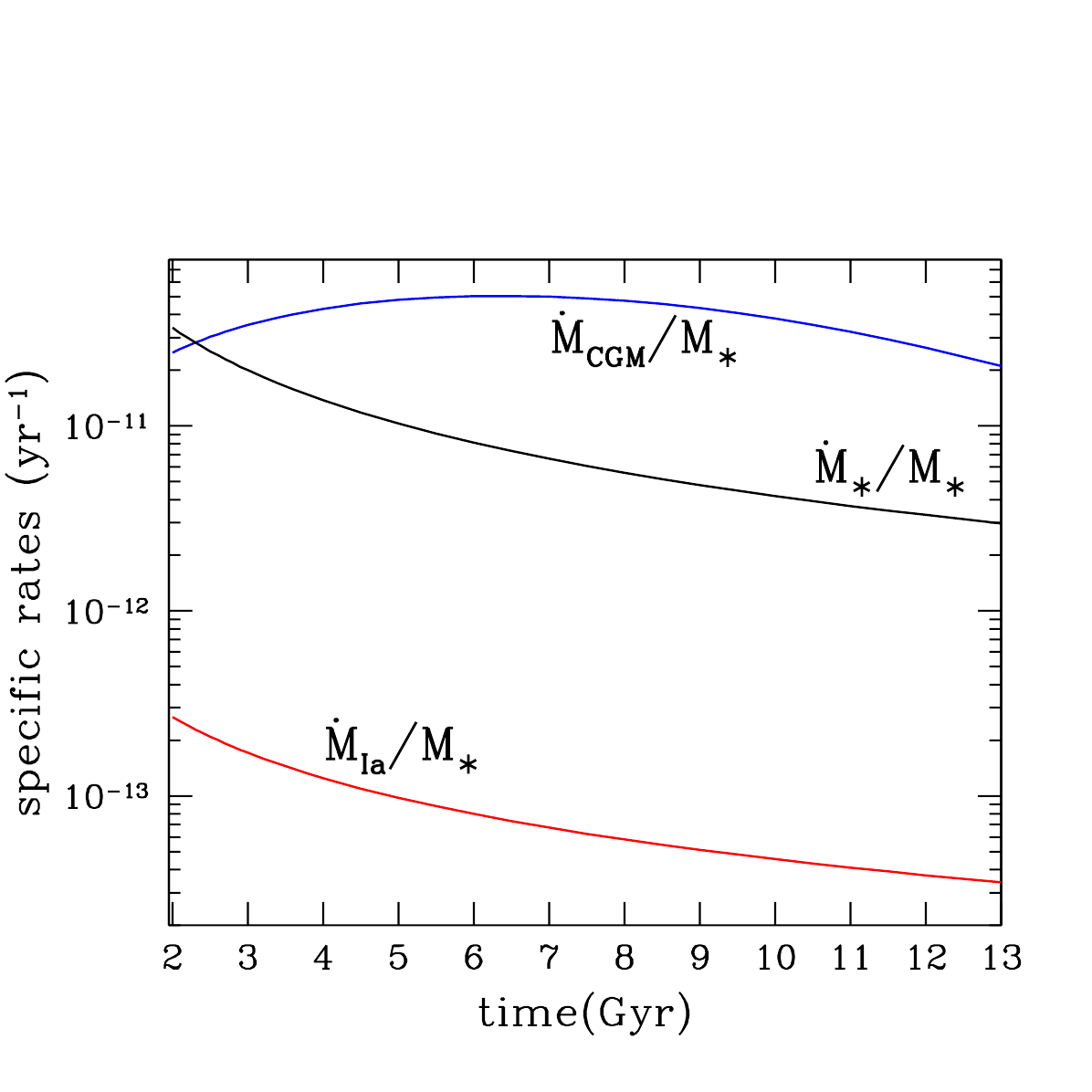}
\vskip -0.4truecm
    \caption{The rates of mass input to the ISM described in Section 2.2, each
normalized to $\Mst$. The aging stellar population inputs are $\dMst$ (in black) and
$\dot M_{\rm Ia}$ (in red); the CGM infall $\dot M_{\rm CGM}$ (in blue) is
parameterized as in C22.}
\label{fig1}
\end{figure}

The mass source terms for the gas flows are given by mass losses from evolved stars, SNIa explosions, and SNII from the new stars formed (see
Appendix B in G19a; Pellegrini 2012; Ciotti \& Ostriker 2012), and by cosmological accretion from the CGM.
Stellar mass losses inject gas with a time decreasing rate $\dot\rho=(\dMst/\Mst) \rhost$, where $\rho$ is the gas density (see Figure \ref{fig1}). 
Over $\approx 10$ Gyr, this mass injection term sums up to a total gas mass of $\simeq 0.1\Mst$.
Figure \ref{fig1} also shows the mass input rate from SNIa explosions ($\dot M_{\rm Ia}$). Following G19a, the time-dependent rate of mass accretion from the CGM is
$\dot M_{\rm CGM}\propto t\,{\rm e}^{-t^2/t_0^2}$,  that approximates the results of cosmological zoom-in simulations for massive ellipticals; C22 adopted $t_0=9$ Gyr,
and the proportionality constant in the formula such that the 
total accreted mass from the CGM between 2 and 12 Gyr  is  $\simeq 0.4\Mst$.
The mass accretion from the CGM is imposed at the outer boundary ($250$ kpc) of the numerical grid, and the CGM mass flux
is weighted by a $\sin^2\theta = R^2/r^2$ angular dependence, so that most of the CGM is injected near the equatorial plane.

\vspace{5truemm}

The various source terms inject into the galaxy also momentum, and internal and kinetic energy
(Section 3 of C22). In particular, the stellar kinematical properties enter 
the thermalization term in the energy equation according to Equation (17) in C22, and the momentum equation through Equation (18)
in C22. In the rotating models, the momentum injection leads to the formation of the cold gaseous equatorial disk, a place
of star formation. The CGM inflow also injects momentum and energy: the CGM injection velocity is half of the free-fall velocity from infinity
(Equation (19) in C22), and the internal energy of the infalling gas is such that its sound
velocity equals the injection velocity (G19a).

Star formation in the cold gaseous disk, that forms in rotating models, is implemented as a result of Toomre instability plus physically
based conditions on gas density and temperature, as described in Equations (20)-(21) in C22.  Star formation is also allowed
to occur everywhere in the galaxy, provided that 1) the gas temperature falls below $4\times 10^4$ K, and 2) the gas number density is higher
than $10^5$ cm$^{-3}$.
Under such conditions the timescale of star formation is given by $\max(\tau_{\rm cool},\tau_{\rm dyn})$, with $\tau_{\rm cool}$ the standard
cooling time, and $\tau_{\rm dyn}$ defined in Equations (23) and (23)  in G19a.
The adopted IMF is such that $\simeq 60\%$ of the mass of newly formed stars is in stars with mass $>8\Msol$ that explode as SNII
on a timescale of $\approx 2\times 10^7$ yrs and inject their mass in the ISM; the motivation for this IMF is discussed in C22.
 
Finally, the implementation of AGN feedback in its radiative  and mechanical (momentum and kinetic energy) components, where the latter
is due to AGN winds, is described in Section 2.7 in G19a, with the small modifications illustrated in Section 3 of C22.
AGN feedback is self-consistently triggered by accretion of low angular momentum gas, along the polar direction, and of
recurrent discharges of gas on the central SMBH due to the Toomre instability in the cold rotating disk.  The implementation of this
second accretion channel is done via the modeling described in Section 3 of C22.
   
The Eulerian hydrodynamical equations are solved with the high resolution grid code MACER, based on the Athena++ code (Stone et al. 2020),
in spherical coordinates $(r,\theta)$, assuming axisymmetry.
More details on the code and the numerical scheme of integration are given in Section 3.1 of C22.
The innermost grid point is placed at 25 pc from the center, the  outermost at 250 kpc.
The age of the galaxy at the beginning of the simulation is 2 Gyr, so that the initial phases of galaxy formation are
terminated, and the flow evolution is followed for 11 Gyr; in the central regions, during outbursts, the fluctuations are followed with
a temporal resolution as short as $\simeq 10^3$ yr.

\renewcommand\arraystretch{1.4}  
\begin{table}
\vspace{2mm}
\centering 
\caption{X-ray model properties at a galaxy age of 10 Gyr}
\vspace{2mm}
\begin{threeparttable}
  \begin{tabularx}{250pt}{c@{\extracolsep{\fill}} c c c c c c  }    
  \toprule 
    Model name   &  $\Lx$   & $\avTf$ &  $\avTx$\\
                 & $(10^{40}$erg s$^{-1}$) & $(10^7\, {\rm K})$ & $(10^7\, {\rm K})$  \\
  (1)            &                   (2)                      &        (3)                &    (4) \\
   \midrule                                                 
  LM$_0$ & 8.09 & 0.59 & 0.59 \\  
  LM$_k$ & 3.06 & 0.76 & 0.60 \\
  LM$_1$ & 2.24 & 0.88 & 0.61 \\
  \midrule 
  MM$_0$ & 44.9 & 1.09 & 0.78 \\
  MM$_k$ & 14.6 & 0.99 & 0.76 \\
  MM$_1$ & 10.1 & 1.01 & 0.74 \\
  \midrule 
  HM$_0$ & 243  & 1.31 & 0.97 \\
  HM$_k$ & 80.2 & 1.18 & 0.96 \\
  HM$_1$ & 55.0 & 1.20 & 0.96 \\
\midrule
  HM$_k^{{\rm new}}$ $\,\,\,\,\,\,\,$ & 57.0 & 1.24 & 0.98 \\
  HM$_k^{{\rm noCGM}}$ & 3.22 & 0.92 & 0.92 \\
  \bottomrule 
\end{tabularx}
\vspace{1.5mm}
\begin{tablenotes}[para,flushleft]
  \footnotesize Notes: (1) model names: for each model mass (LM, MM, HM) the subscript indicates the type of
  azimuthal stellar motions, described in Section 2.1;  in order of increasing importance of the rotational support,
  $0$ means no ordered rotation ($k=0$), $k$ indicates the exponentially  declining rotation  $\ke(r)$ in Equation (\ref{eq:ke}), and $1$ the isotropic rotator ($k=1$).
    (2) the total luminosity $\Lx$ of the hot gas in the $0.3-8$ keV energy band.
    (3) the $0.3-8$ keV average temperature, computed as detailed in Appendix A3, for an aperture of $5\Reff$.
    (4) the $0.3-8$ keV average temperature computed for the whole galaxy.\\
    The first nine models were run by C22, the last two have been run for the discussion in Section~\ref{profiles}.
\end{tablenotes}
\end{threeparttable}
\label{tab2}
\end{table}
\renewcommand\arraystretch{1.}
\vspace{5truemm}

\begin{figure*}
\vskip -3truecm
\hskip -1truecm  
\includegraphics[width=0.77\linewidth, keepaspectratio]{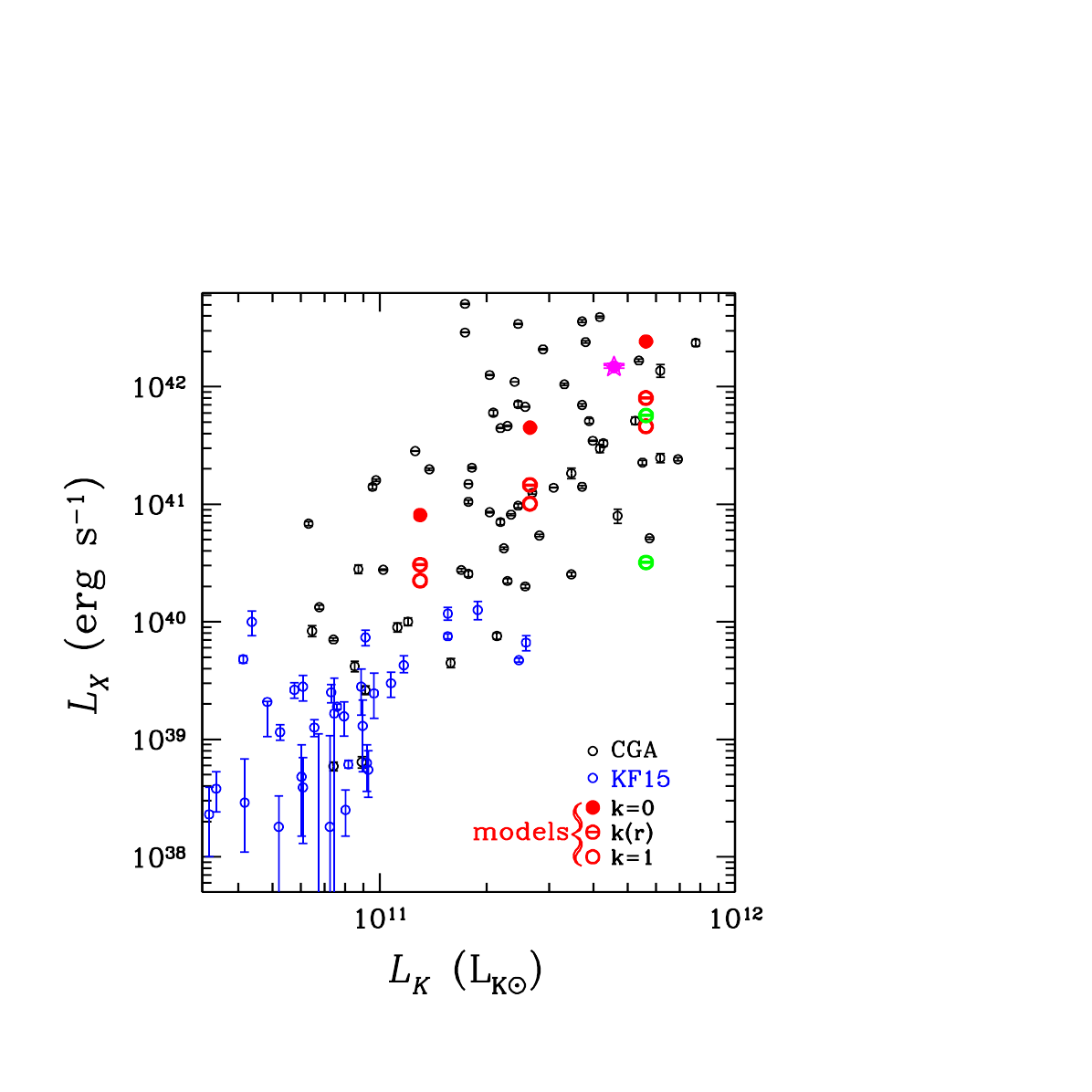}
\hskip -4.2truecm
\includegraphics[width=0.77\linewidth, keepaspectratio]{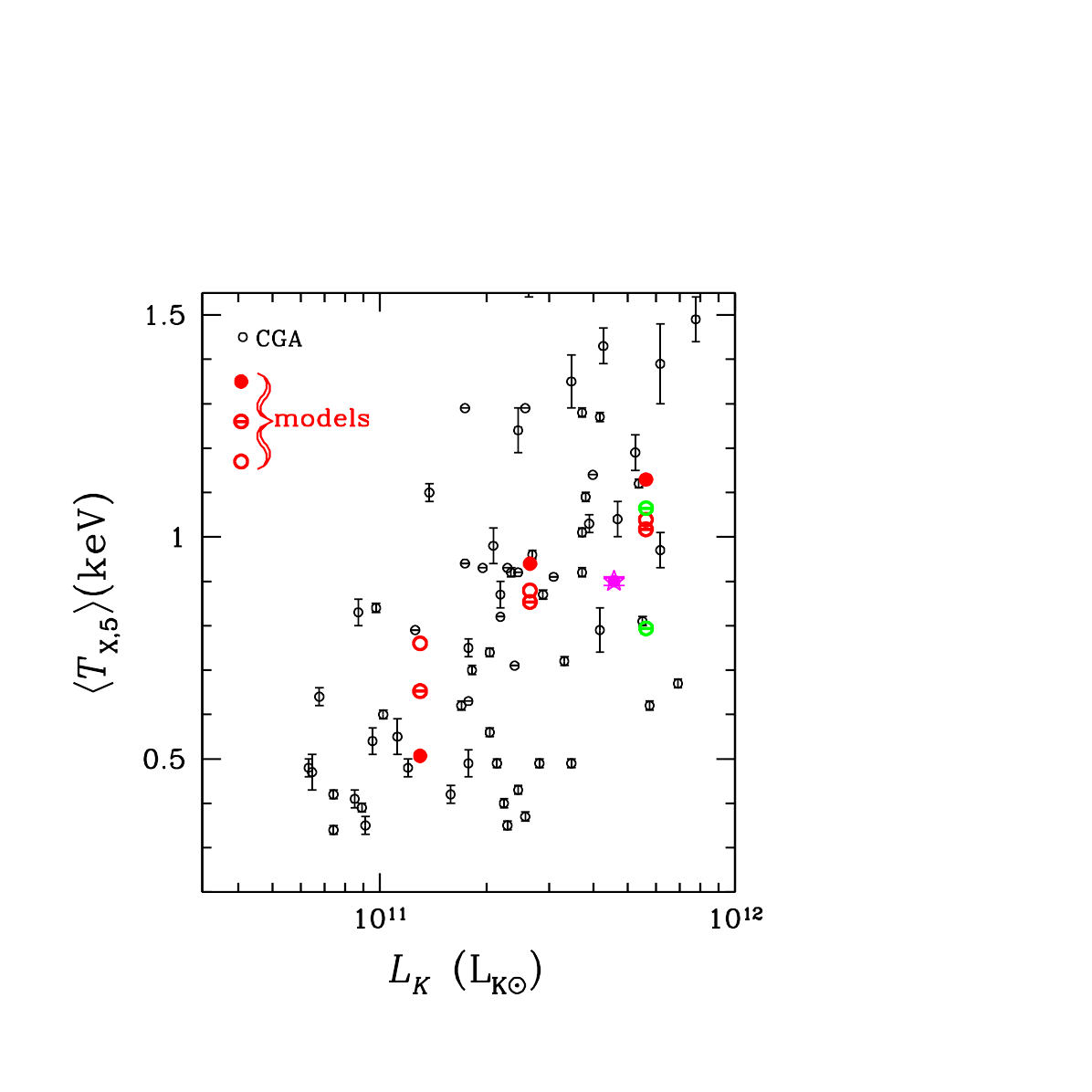}
\vskip -3.8truecm
\hskip -1truecm
\includegraphics[width=0.77\linewidth, keepaspectratio]{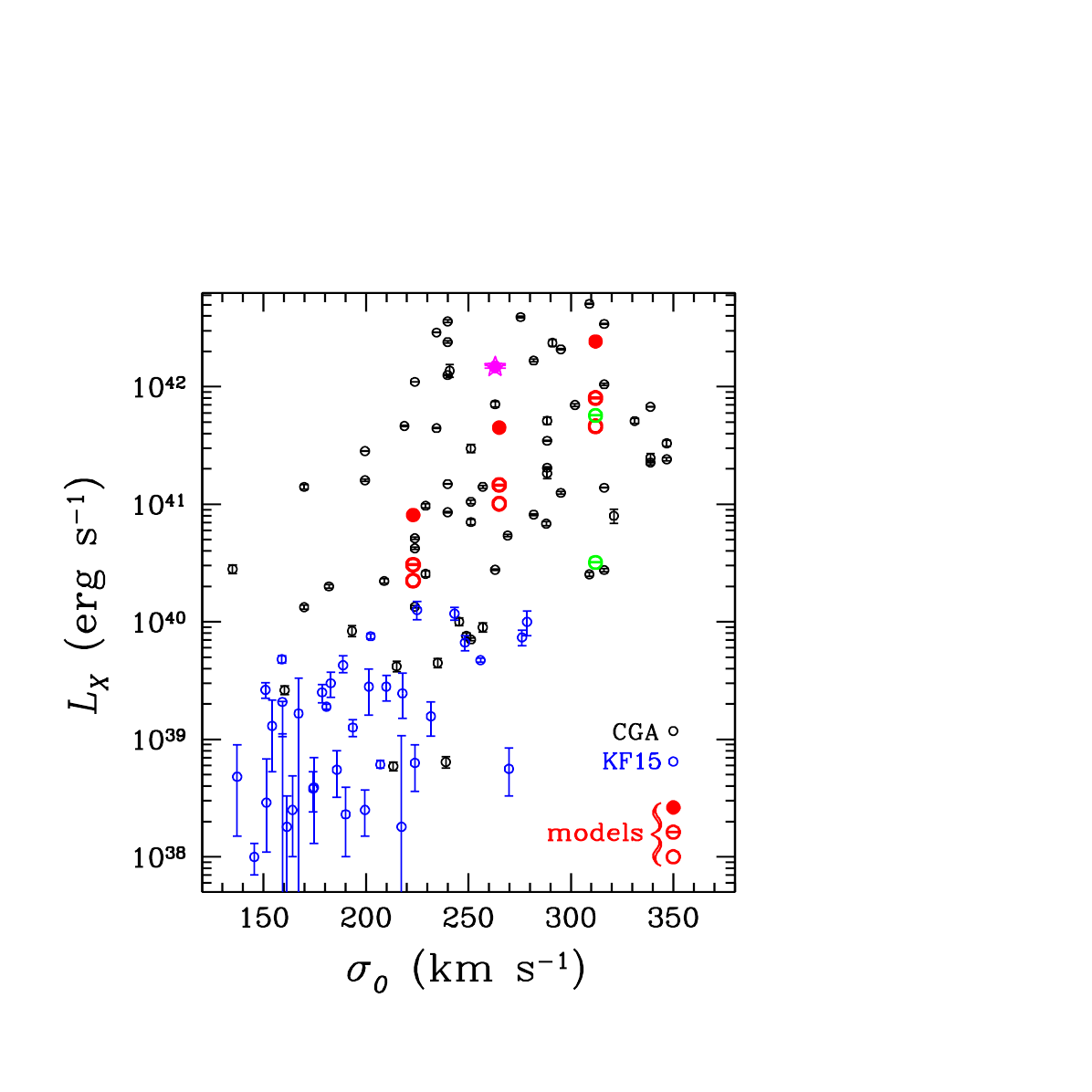}
\hskip -4.2truecm
\includegraphics[width=0.77\linewidth, keepaspectratio]{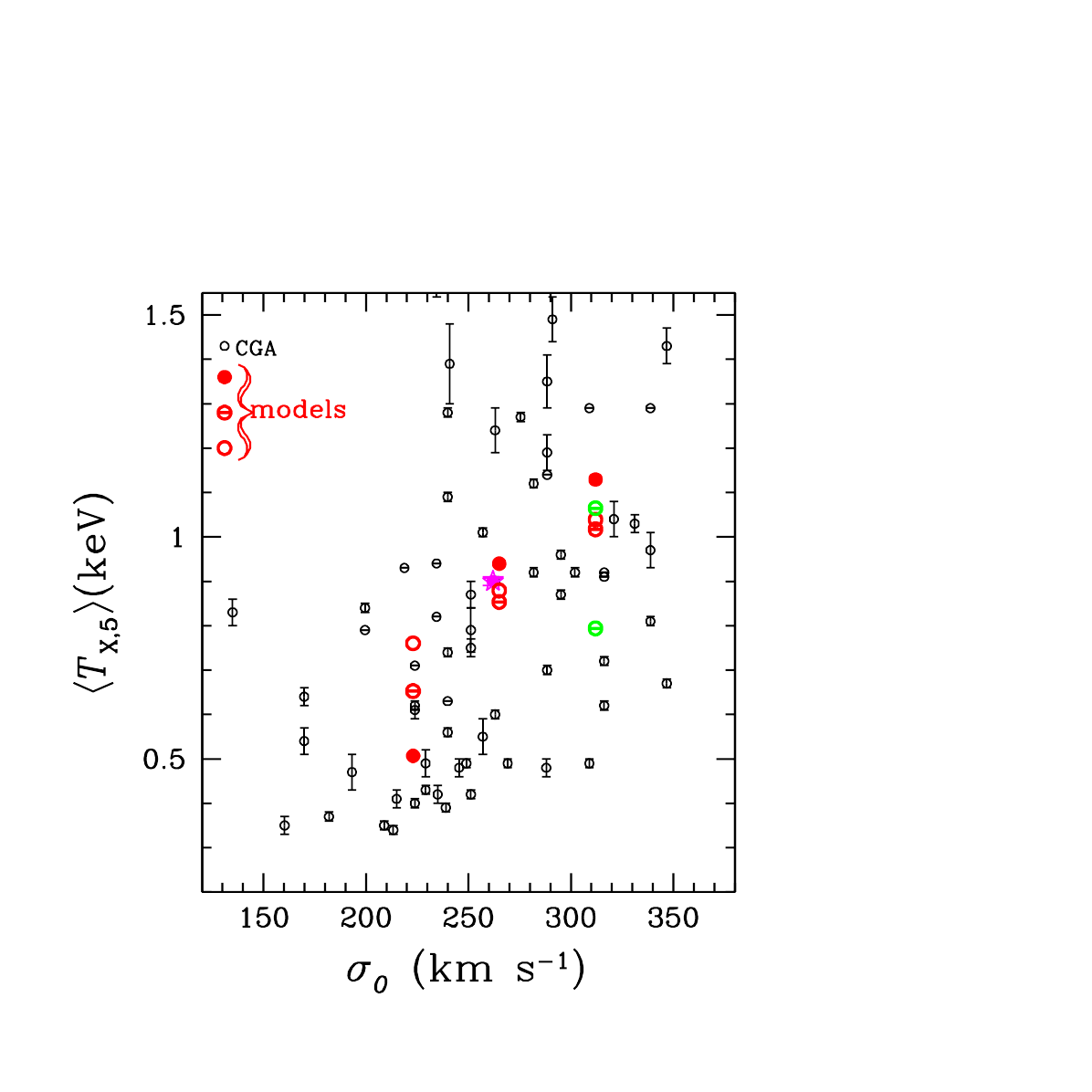}
\vskip -1truecm
\caption{X-ray properties of the models in Table~\ref{tab2} compared with the corresponding properties for two samples: the ETGs in
  the $Chandra$ Galaxy Atlas (K19; CGA in the legend) and those in the ATLAS$^{\rm 3D}$ sample observed with $Chandra$
(Kim \& Fabbiano 2015; KF15 in the legend). Upper left: the total $0.3-8$ keV $\Lx$ versus the K-band galactic luminosity $\Lk$.
  Upper right: the $0.3-8$ keV average temperature within $5\Reff$, $\avTf$, for the models and CGA galaxies, versus $\Lk$.
  Lower left: $\Lx$ vs. the central stellar velocity dispersion $\sigma_0$; for observed galaxies, $\sigma_0$
  comes from Kim \& Fabbiano (2015), K20, Babyk et  al. (2018); for the models, $\sigma_0$ is $\sigmas(0)$ in Table~\ref{models}.  
  Lower right: the same $\avTf$ as in the upper right panel versus $\sigma_0$. See Section~\ref{LxTx} for more details. In all panels, the pink
  star shows the galaxy NGC5129, and the two green symbols show the two additional models HM$_k^{{\rm new}}$ and HM$_k^{{\rm noCGM}}$
  in Table~\ref{tab2},   discussed in Section~\ref{profiles}.   }
\label{fig2}
\end{figure*}
\vspace{5truemm}

\section{X-ray luminosities and temperatures of the hot gas}\label{LxTx}

We present here a first test for the viability of the implementation of the input physics in the models: the agreement (or not) with observed values
of their global X-ray properties, as the hot gas luminosity $\Lx$ and its average temperature $\avTx$.
For this test we look at the distribution of the models in  diagnostic planes as $\Lx -\Lk$ and $\Lx - \avTx$,  where
$\Lk$ is the K-band galaxy luminosity.
To determine model quantities analog of those measured in the X-rays, we proceeded as follows (more details in Appendix A).
The  X-ray emissivity in the $0.3-8$ keV energy band, considering the possibility of absorption by intervening cold gas within the galaxy (e.g., due to the cold disk),
was integrated along the line of sight, for an edge-on view of the galaxy; the result was a surface brightness map $\SigX$, that was integrated in the image plane to
compute the total $\Lx$ and that within a cylinder of radius equal to 5$\Reff$ and axis along the line of sight, $\Lxf$ (Equation A10);
the circularized surface brightness profile $\SigX(R)$ was determined as an angle averaged quantity over an annulus centered at $R$ (Equation A12).
Average temperatures were derived first performing a projection along the line of sight of the gas temperature weighted with the
X-ray emissivity,  including again the possibility of intrinsic absorption (Equation A11);  from the temperature map so obtained, we evaluated the circularized
temperature profile $\Tx(R)$ (Equation A12); weighting $\Tx(R)$ with $\SigX(R)$, we computed the average temperature $\avTx(R)$ within radius $R$ in the image
plane\footnote{The average temperatures and the $\Tx(R)$ profiles obtained with this procedure are emission-weighted quantities, as 
  are, with good approximation, the observed temperatures used for comparison in this work; see K19, Truong et al. (2020).}
(Equation A13). In the following we consider the average temperatures $\avTf$, within an aperture of radius 5$\Reff$, and $\avTx$, for the whole galactic image.

Figure~\ref{fig2}  shows the position of the models in the $\Lx -\Lk$, $\Lx -\sigma_0$ (the central stellar velocity dispersion), $\avTf - \Lk$ and $\avTf - \sigma_0$
planes. Red symbols indicate the nine models in C22, and green symbols two additional models discussed in Section~\ref{profiles}, at a representative galaxy age
of 10 Gyr (Table~\ref{tab2})\footnote{C22 report $\Lx$ from a spherical volume of $r<5\Reff$, at an age of 13.7 Gyr, in their Table 2; also, there,
  $\Lx$ of model HM$_0$ should read 129 instead of the reported 12.9,  due to a typo.}. Each model can be identified from its$\Lk$ (in Table~\ref{models})
its $\Lx$ (in Table~\ref{tab2}), and its rotational properties that are also specified in the figure. Observed quantities in Figure~\ref{fig2} derive from $Chandra$ pointings:
$\Lx$ is the 0.3--8 keV hot gas luminosity, measured from within the largest available radius (Kim \& Fabbiano 2015, K19); $\avTf$ is  
the average luminosity-weighted temperature within a circle of radius 5$\Reff$ or, when not available (13 cases),  within a smaller radius (from K19).
In all four panels,  the distribution of the models falls within that of observed ETGs, and also reproduces the general observed trends.
In the top left panel ($\Lx -\Lk$),  the model $\Lx$ at each $\Lk$ decreases for an increasing amount of ordered rotation; this confirms previous findings, obtained
also with different galactic structures and different codes (i.e., Negri et al. 2014a,b). The trend is explained by
the tendency of rotating flows to induce gas cooling in the central regions; in addition, more gas mass can be ejected as the gas centrifugal support increases, thus 
the overall effect of rotation is to produce more cold gas and less hot ISM (see Table 2 in C22;  Posacki et al. 2013).
A trend in this sense has been also observed: flatter galaxies, that tend to rotate more, show on average lower $\Lx$ (Eskridge et al. 1995,
Sarzi et al. 2013, Juranova et al. 2020). In the bottom left panel ($\Lx -\sigma_0$),
$\sigmas(0)$ in Table~\ref{models} is used as a proxy for the projected $\sigma_0$ of the models; these fall within the observed distribution, 
and follow its general trend. We note that this plot is not just a replication of the $\Lx -\Lk$ plane, because even though $\Lk$ and $\sigma_0$
correlate through the Faber-Jackson relation, they do so with a large scatter.

The right panels in Figure~\ref{fig2}  show the $\avTf$ vs. $\Lk$ and $\avTf$ vs. $\sigma_0$ planes, for the set of ETGs with $\avTf$ available in the $Chandra$ Galaxy Atlas
(K19). For these panels we adopted the temperature within an aperture of $5\Reff$, instead of the global $\avTx$, because a temperature
averaged over a smaller region is more sensitive to the model properties, as the rotational support; instead, $\avTx$
depends almost exclusively on the galaxy mass, and is very similar for models of the same mass (see Table~\ref{tab2}).
At variance with what happens for $\Lx$, that decreases for increasing rotation at each $\Lk$ (and $\sigma_0$), here
the relation between rotation and $\avTf$ depends on the galaxy mass: in MM and HM models, $\avTf$ is lower when rotation is present, because the cold
disk formation leaves a lower hot gas density in the central (typically hotter) region (see also Section~\ref{profiles}); in LM models,
instead, rotation is more effective in favouring the development of winds, that are hotter than inflowing gas, and thus the opposite trend of 
$\avTf$ with rotation establishes (see also Negri et al. 2014b).

\begin{figure*}
 \vskip -3.truecm 
  \hskip -1truecm 
  \includegraphics[width=0.75\linewidth, keepaspectratio]{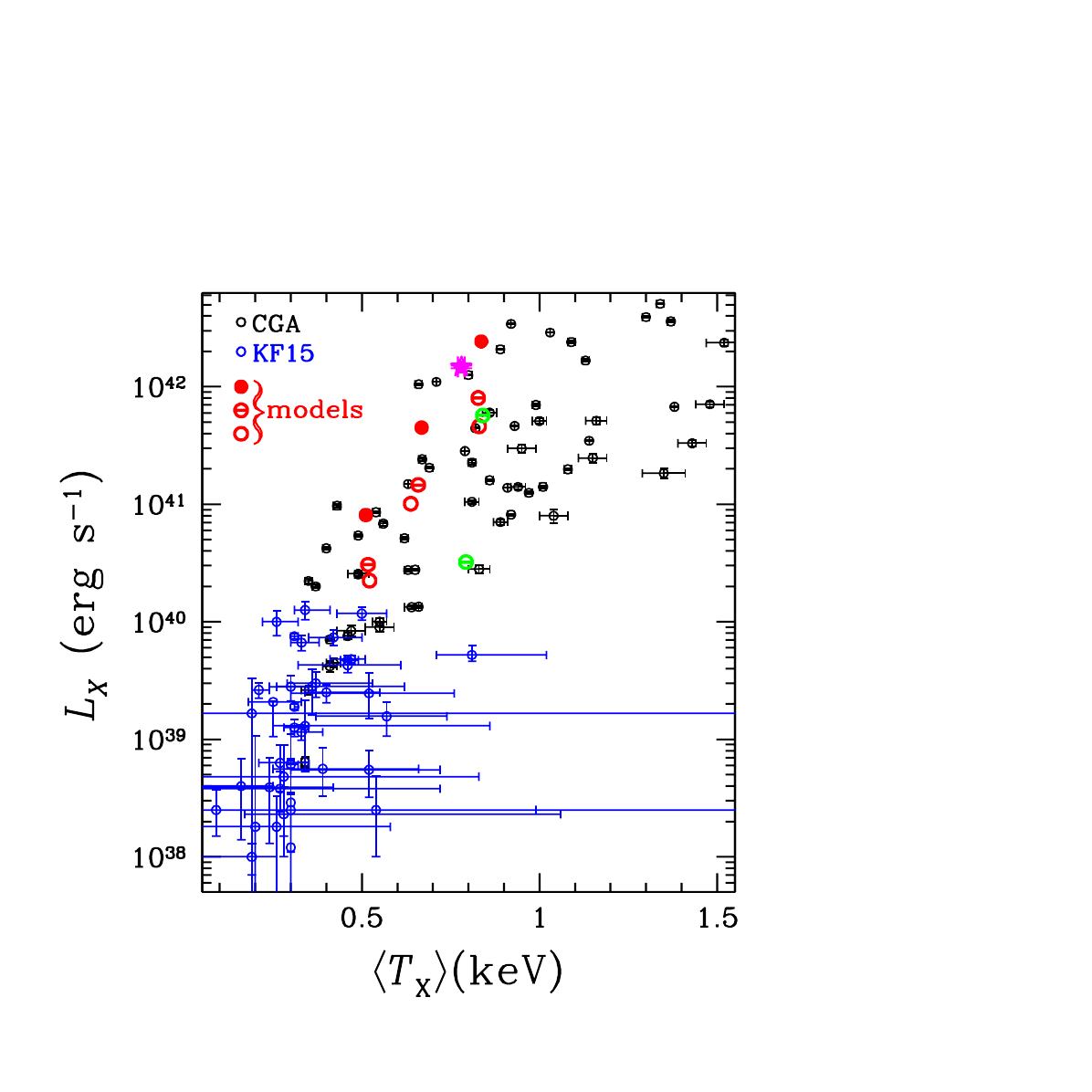}
  \hskip -4.truecm 
  \includegraphics[width=0.75\linewidth, keepaspectratio]{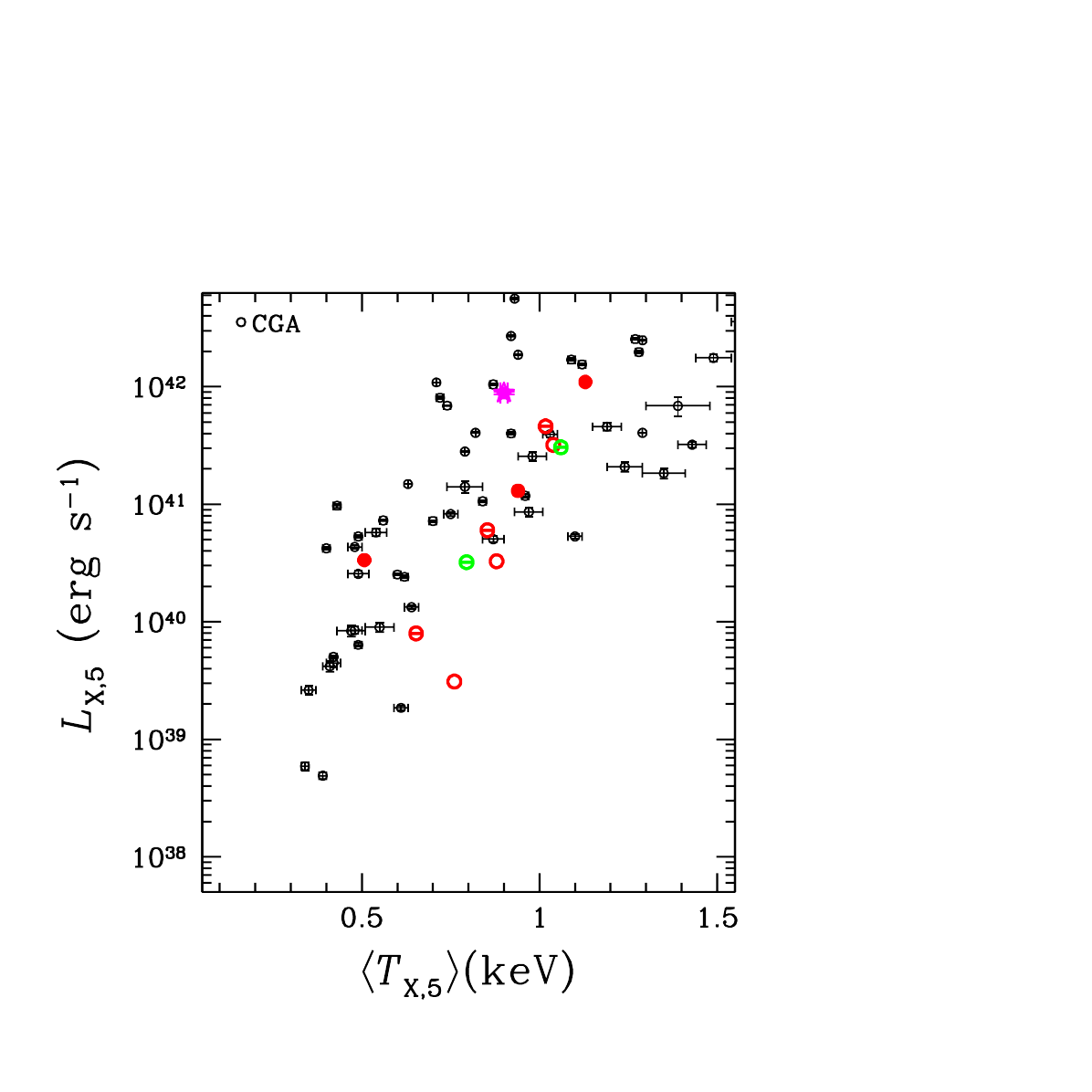}
  \vskip -1truecm 
  \caption{Left panel: $\Lx$ versus $\avTx$ for the models in Table~\ref{tab2}, and observed ETGs; references for the observed quantities, symbols and meaning
    of the legend are the same as in Figure~\ref{fig2}, upper left panel. Right panel:  $\Lxf$ versus $\avTf$ for the models in Table~\ref{tab2},  and observed ETGs, from K19 only.
  The pink star shows NGC5129, and the two green symbols the two additional models HM$_k^{{\rm new}}$ and HM$_k^{{\rm noCGM}}$   discussed in Section~\ref{profiles}.    }
\label{fig3}
\end{figure*}

Another common diagnostic diagram is the $\Lx-\avTx$ plane (e.g., Kim \& Fabbiano 2015, Goulding et al. 2016, Babyk et al. 2018),
shown here in Figure~\ref{fig3}, left panel; as for $\Lx$, also the observed  $\avTx$  derives from the largest extraction radius available.
In this plane the positions of the models fall within those of observed ETGs; each family of LM, MM and HM models is located along an almost vertical
column of red (and green) points, since $\avTx$ depends mostly on the galaxy mass.
Except for one green point, however, the models tend to reside in the region of the more X-ray
luminous ETGs, at a fixed $\avTx$, or of the lower  $\avTx$, at a fixed $\Lx$. To investigate further this point, we made 
closer the comparison between models and observations, by plotting strictly matching quantities in terms of the extraction region for
the computation of luminosity and average temperature. 
The right panel of Figure~\ref{fig3} thus shows $\Lxf$ versus $\avTf$; here only those ETGs on the left for which these quantities are available
in the $Chandra$ Galaxy Atlas (K19) are plotted. In the right panel the models have moved towards larger temperatures ($\avTf > \avTx$ for most of them,
Table~\ref{tab2}), while the distribution of ETGs overall has not changed much (for a number of them, $\avTx$ in the left panel is already estimated at
or close to 5$\Reff$). Also, $\avTf$ is more different than $\avTx$ for models of the same mass\footnote{The difference between $\avTf$ and $\avTx$, as that
  between $\Lxf$ and $\Lx$, depends of course on the shape of $\SigX(R)$ and $\Tx(R)$ (Equation A13); for example, the difference is
  lower for a more peaked $\SigX(R)$, that reduces the importance of the galaxy regions outside 5$\Reff$.}, so the models' positions are more spread over the plane.
The result is that now the models fall within, and cover well, the range of observed values.
We note finally that the green point with the  lower $\Lx$ in Figure~\ref{fig3} (and Figure~\ref{fig2} as well)  is a model in all equal to 
HM$_k$ but evolved without CGM accretion (model HM$_k^{{\rm noCGM}}$ in Table~\ref{tab2}). Its $\Lx$ is much lower than that of HM$_k$, indicating how this kind of accretion
can produce a large variation in $\Lx$, at fixed $\Lk$. Its $\avTx$ is equal to $\avTf$, due to its peaked  $\SigX(R)$.
This model will be considered further in Section~\ref{profiles}.

In conclusion, $\Lx$, $\Lxf$, $\avTx$, and $\avTf$ of the models are found within the observed range, and also their trend with $\Lk$ and $\sigma_0$
is satisfactory: more massive galaxies are more X-ray luminous and
hotter than less massive systems, a well know manifestation of the larger binding energy per unit gas mass in larger galaxies (as indicated, e.g.,
by the Faber-Jackson relation). Moreover, less rapidly rotating systems are more X-ray luminous than more rotating ones of the same mass,
a trend also possibly present in the observations. A tendency for the models to occupy the upper envelope of the observed $\Lx$ distribution, at fixed $\avTx$,
disappears when considering $\Lxf$ and $\avTf$.

\begin{figure}
  \hskip -1truecm
  \includegraphics[width=1.4\linewidth,  keepaspectratio]{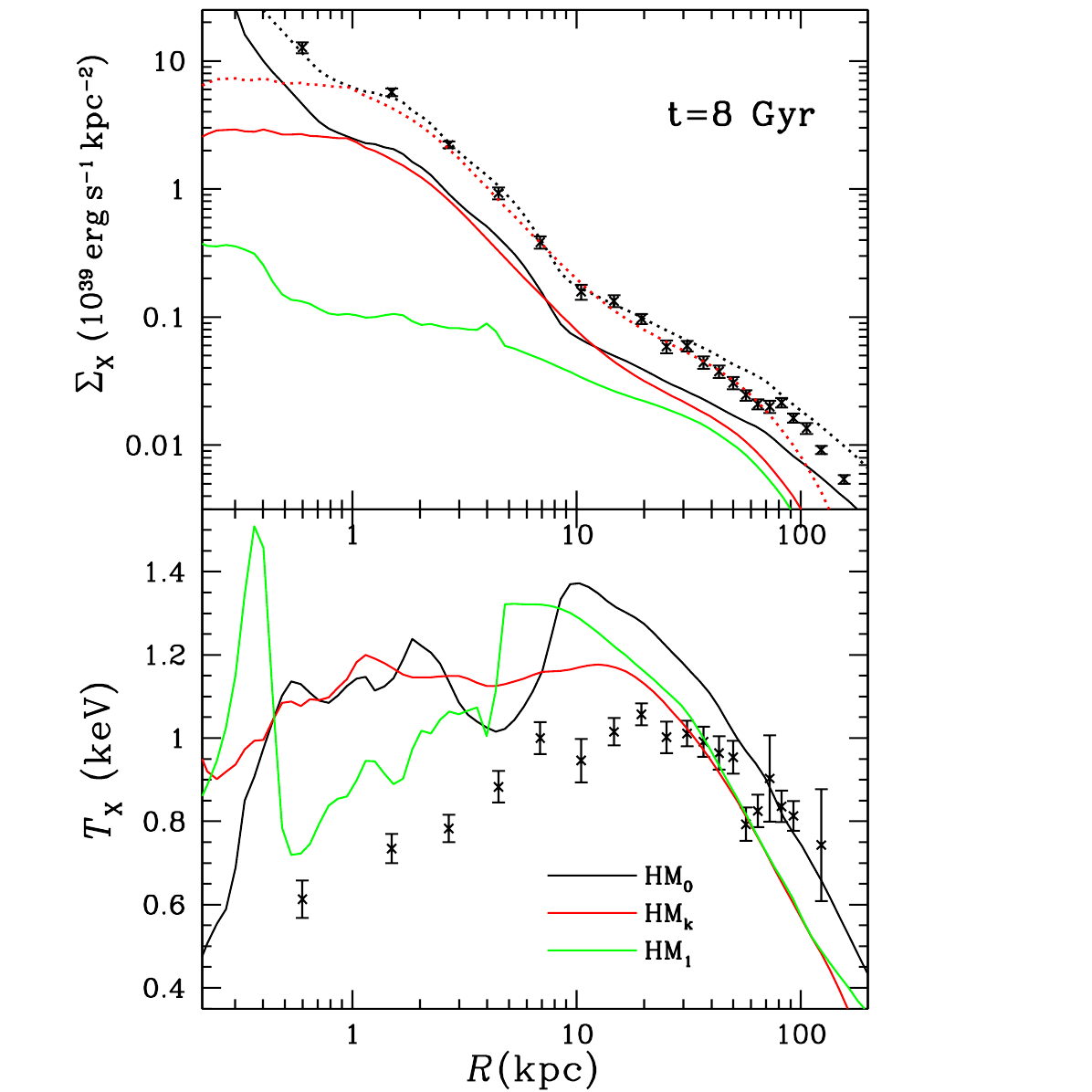}
  \caption{The $0.3-8$ keV circularized surface brightness profiles $\SigX(R)$ (upper panel), and temperature profiles
    $\Tx(R)$ (lower panel), at an age of
    8 Gyr, for the three HM models of C22 (Table~\ref{tab2}) shown with solid lines of different colors (as specified in the legend of the lower
    panel); with the same colors, the dotted lines show the brightness profiles of the HM$_0$ and HM$_k$ models scaled-up by a factor of 2.5 (Section~\ref{profiles}).
    The corresponding profiles for  NGC5129 are shown by black symbols with errorbars (from K19).}
  \label{fig4}
\end{figure}

\section{X-ray surface brightness and temperature profiles}\label{profiles}

Here we explore how the brightness profile $\SigX(R)$ and the temperature profile $\Tx(R)$ of the models
compare with those observed; for this purpose,  these profiles were computed in a way
to obtain quantities analog  to those measured (Section~\ref{LxTx}; Appendix A).
Indeed, global X-ray properties even consistent with observations could be associated with $\SigX(R)$ and $\Tx(R)$
different from those observed; thus this study can provide additional information on the performance
of the models and on the possible need for modifications in the input physics. Since the simulations of C22 were not designed to reproduce
a specific ETG, we first select a representative galaxy in the $Chandra$ Galaxy Atlas and carry out a comparison
with its $\SigX(R)$ and $\Tx(R)$, recalling that the analysis can  sometimes only be qualitative.
Next, in Section~\ref{cga}, we extend the comparison to more ETGs in this Atlas.
  
As a representative galaxy we selected NGC5129, an X-ray bright ETG with an extended hot halo.
Its $\Tx(R)$ profile fits in the universal shape and in particular is 
the prototypical example of the most commonly observed type of profiles, the HB one (K20; see Section~\ref{Intro}).
At the distance of 103 Mpc, NGC5129 is a moderately rotating  E3-E4 galaxy, with $\Lk = 4.6\times 10^{11}L_{\odot,K}$, a stellar mass of $\Mst = 7.2\times
10^{11}M_{\odot}$, and $\Reff= 14$ kpc (Veale et al. 2017), all properties that make it similar to the galaxies of the HM family.
NGC5129 is also the dominant galaxy in a poor galaxy group, and its estimated age within $R_{\rm e}/8$ is 7.4 Gyr (Gu et al. 2022). Its $\Lx$, $\Lxf$, $\avTf$ and $\avTx$ are
shown in Figures~\ref{fig2} and~\ref{fig3} with the pink star. NGC5129 was also studied in the X-rays by Eckmiller et al. (2011), Bharadwaj et al. (2014), and Nugent et al. (2020).
All these studies found a central positive gradient in the temperature profile, a peak of $kT\simeq 1.1$ keV at $R\simeq 20$ kpc,
and then a decline out to $R\approx 200 $ kpc. The most spatially detailed $\SigX(R)$ and $\Tx(R)$ profiles are those determined from $Chandra$
data by K19, and we consider them in the following.

 \begin{figure*}
\hskip -1truecm  
  \includegraphics[width=0.65\linewidth,  keepaspectratio]{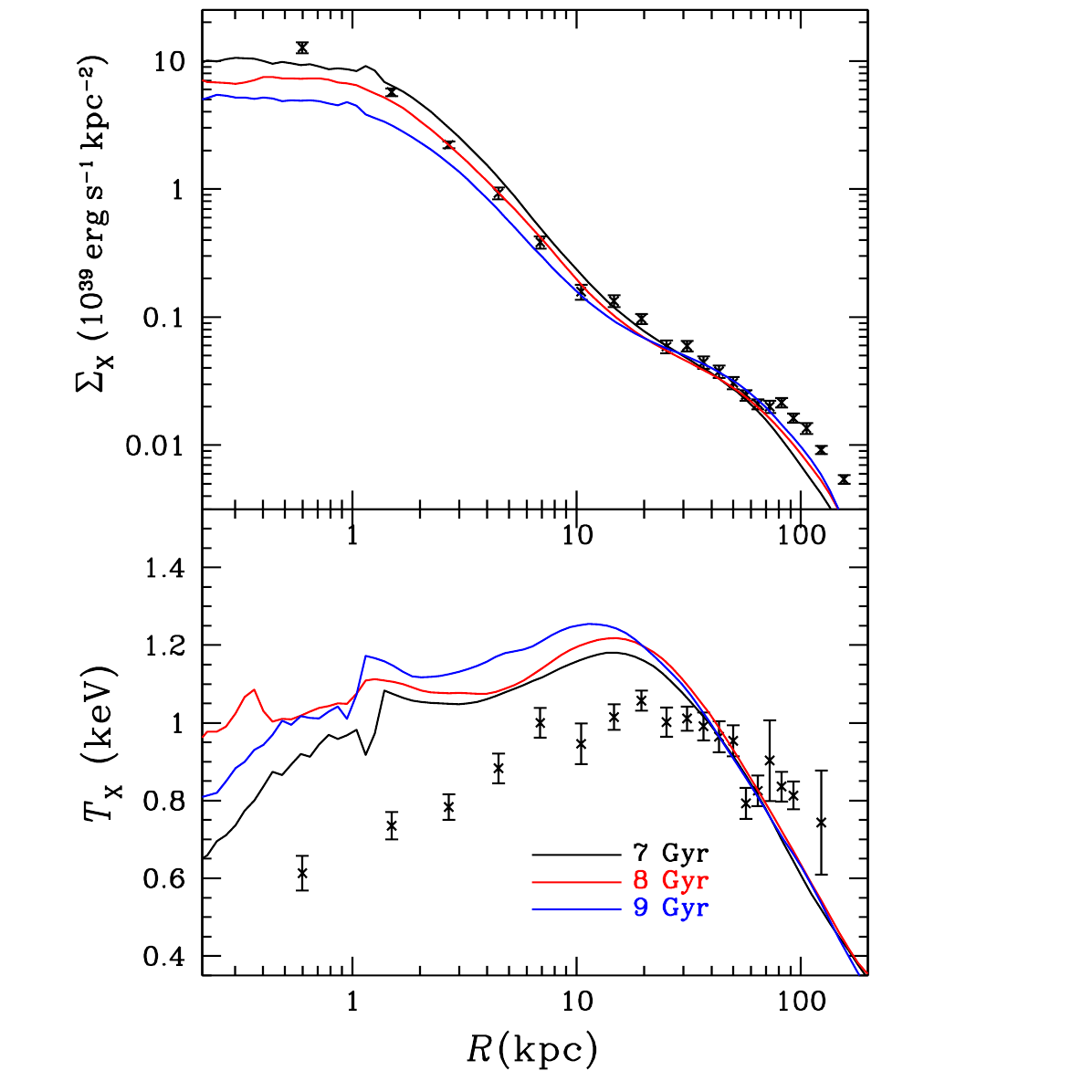}
 \hskip -2truecm 
\includegraphics[width=0.65\linewidth, keepaspectratio]{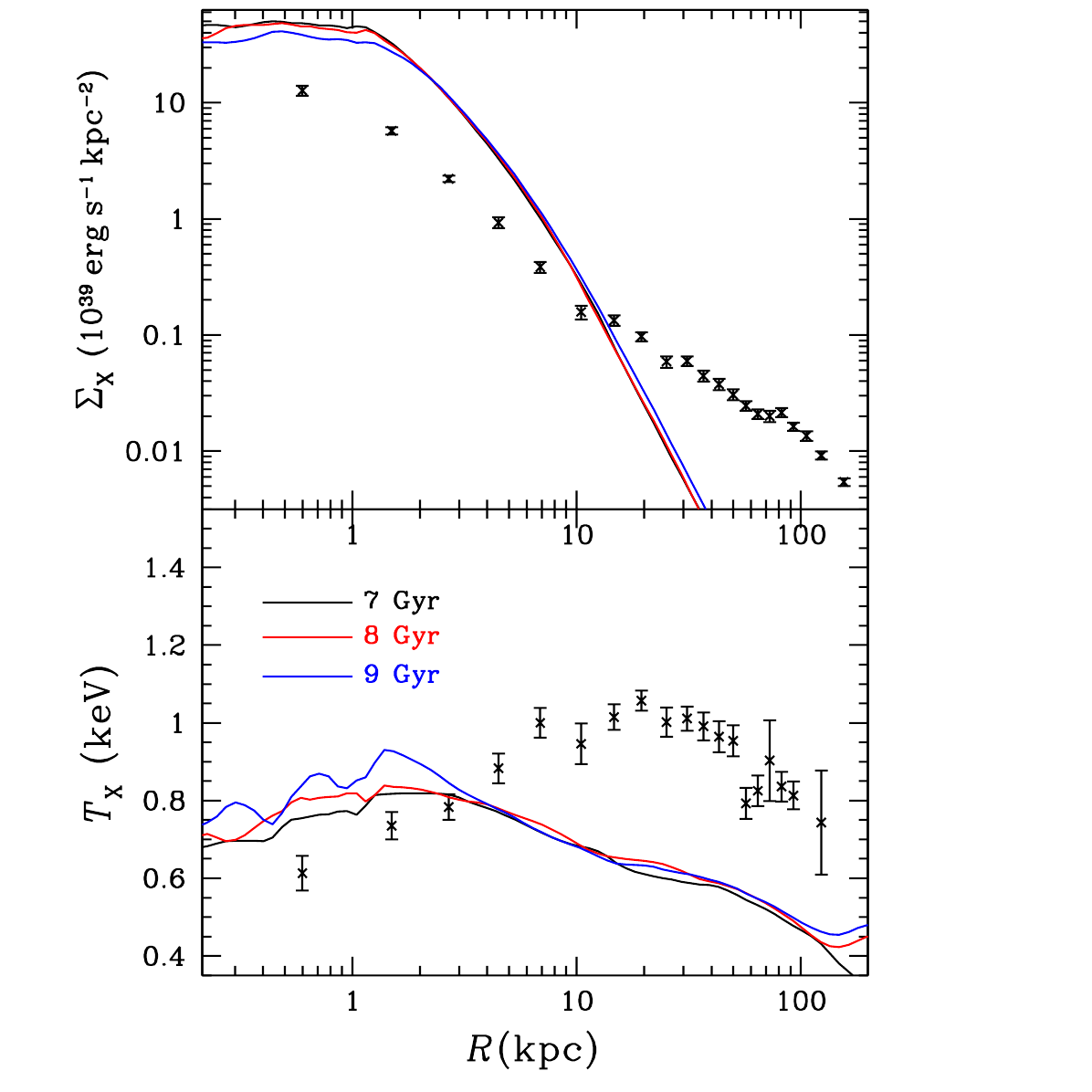}
\caption{ Left panels: model HM$_k^{{\rm new}}$ at three representative epochs, compared with NGC5129 (points with errorbars, from K19).
The model $\SigX(R)$ have been scaled-up by a factor of $\simeq 3$  to match the 
  profile of NGC5129  (Section~\ref{profiles}); notice how the model profiles reproduce the change of slope outer of $R\simeq 10$ kpc. 
  Right panels: model HM$_k^{{\rm noCGM}}$, at the same epochs on the left, compared with NGC5129; this model is in all equal to the HM$_k$ models, but has been evolved without CGM mass accretion.
  The model $\SigX(R)$ has been rescaled as for HM$_k^{{\rm new}}$  for the plotting purpose (Section~\ref{profiles}). 
The $\Tx(R)$ profiles of HM$_k^{{\rm noCGM}}$,  in the outer region, are lower than observed, proving that CGM accretion 
  is here fundamental to shape the temperature profile.}
\label{fig5}
\end{figure*}
\vspace{5truemm}

For the three HM models, Figure~\ref{fig4} shows the circularized $\SigX(R)$  and $\Tx(R)$ profiles
together with those of NGC5129; here $R$ is the distance from the galactic center in the X-ray image, and the models are
viewed edge-on (Appendix A.3). The model profiles refer to an age of 8 Gyr.
It is apparent how the $\SigX$ shape becomes more similar to that of NGC5129 when decreasing the amount of the stellar ordered rotation (i.e., going from
the green to the red to the black solid lines); rotation tends to make the brightness profile flat within $R\simeq 10$ kpc, a result in line with what obtained in past
simulations (Brighenti et al. 2009, Negri et al. 2014a). In order to compare more closely $\SigX(R)$ of the non-rotating HM$_0$ model with that of NGC5129,
we scaled its $\SigX(R)$ up  by a factor of $\simeq 2.5$, and  obtained the black dotted line in Figure~\ref{fig4}.
The scaled $\SigX$ shows a good match with that observed; however, it seems to be increasing too steeply at the center ($R\la 500$ pc), and it keeps above
the observed profile for $R\ga 30$ kpc.  The same scaling operation applied to $\SigX(R)$  of the mildly rotating HM$_k$
(red dotted line in Figure~\ref{fig4}) also provides a good match with observations over a radial range from $\simeq 1$ to $\simeq 70$ kpc; outside of this range,
it is lower than observed. The $\SigX(R)$ of the highly rotating HM$_1$ model is too discrepant and no scaling was tried.
We note that the $\SigX$ shapes remain similar during the last Gyrs of evolution, for each of the HM$_0$, HM$_k$ and HM$_1$ models, therefore the choice of the age
is not crucial for the above conclusions. From a physical point of view, the scaling of $\SigX$ by a factor of $\simeq 2.5$ can be produced by a uniform
increase of the gas density by $\simeq 50\%$, a variation that is not unreasonable to hypothesize for NGC5129, 
considering that HM models were not tailored on it.
$\SigX$ instead would not scale similarly for a uniform temperature variation, because the 0.3--8 keV emissivity is weakly dependent on the 
temperature when it varies in the range of  $\simeq 0.3 - 1.2$ keV.

The lower panel of Figure~\ref{fig4} shows the temperature profiles of the three HM models, at the same epoch of 8 Gyr.
In the outer galactic region ($R\ga 30$ kpc), $\Tx(R)$ has the correct shape and range of values, while it is  
different from that observed inside $R\simeq 20$ kpc. In particular, the observed bump is absent, and all models show larger temperatures.
Model HM$_0$  shows the largest disagreement with the NGC5129 temperature profile, being almost everywhere too hot.
Of the two remaining models, HM$_1$ shows unobserved temperature fluctuations, while HM$_k$ seems the least discrepant with observations.
Can modifications of the HM$_k$ model give a temperature profile that better reproduces the observed one, and at the same time
maintain the good agreement of $\SigX(R)$ with that of NGC5129?
In order to investigate this aspect, we explored some changes in various parameters of the input physics, while keeping the same galaxy structure and
rotation properties of HM$_k$. Given the computational time required by the simulations,
a full parameter space exploration is prohibitive. Some experiments involving changes in the AGN feedback parameters (as the AGN wind opening angle
and velocity), or in the energy injected by SNII's from star formation in the central regions, did not produce improvements.
Instead, changes in the implementation of the environmental accretion produced variations in $\Tx(R)$ in the sought directions.
In particular,  this was the case for an increase of the CGM accretion velocity imposed at the outer boundary by a factor of 1.5
(from 0.5 to 0.75 of the galaxy free-fall velocity, see Section 2.2),  with the CGM mass inflow ${\dot M}_{\rm CGM}$  kept the same. 
 The $\Lx$, $\Lxf$, $\avTx$ and $\avTf$ of this variant of HM$_k$ (hereafter HM$_k^{{\rm new}}$) are 
 shown in Figures 2 and 3 by the green points with the larger luminosity and temperature values; reassuringly, they still fall within the observed range.
Figure~\ref{fig5} (left panel) shows $\SigX(R)$  and $\Tx(R)$ of HM$_k^{{\rm new}}$, at three different epochs, close to the age of NGC5129.
Similarly to what done for HM$_0$ and HM$_k$, here $\SigX(R)$ is up-scaled by a factor of $\simeq 3$. The agreement of the rescaled $\SigX(R)$ profiles with the
observed one is still good: the observed shape between $R=1$ and $R=10$ kpc is well reproduced, and also its flattenings inside $R=1$ kpc, and outside $R=10$ kpc.
It is remarkable that the $\Lx$ value resulting for the scaled model is close to that measured for NGC5129 from its
  spectrum, i.e., $\Lx=1.48\times 10^{42}$ erg s$^{-1}$ within $R=145$ kpc  (K19).
The HM$_k^{{\rm new}}$ temperature profile is less spatially fluctuating and shows a  better defined and smoother bump, with respect
to that of HM$_k$; the presence and location of this bump, that extends from  8 kpc to 20 kpc, make $\Tx(R)$ of HM$_k^{{\rm new}}$
closer to the observed one. Outside of $R\simeq 20$ kpc, the decline in $\Tx(R)$ and its values match those observed.
Inside of $R\simeq 20$ kpc, however, the slope of $\Tx (R)$ is now similar to that observed, but the $\Tx(R)$ values remain larger by $\simeq 30\%$.
Note that some residual uncertainties, produced by the use of different emission models and atomic data, might still be present in the measured temperature profile.
We discuss further possible origins of the discrepancy in temperature in Section~\ref{conclu}.

The importance of  CGM accretion in determining the brightness and temperature profiles is especially revealed by an experiment where it was suppressed. This model
(HM$_k^{{\rm noCGM}}$) is shown in Figures~\ref{fig2} and~\ref{fig3} by the green points with the lower luminosities and temperatures.
$\Lx$, $\Lxf$, $\avTf$ and $\avTx$ are still within the observed range, though on the lower side of the distribution of values.
$\Lx$ of HM$_k^{{\rm noCGM}}$ is much lower than that of the other HM models, a combined consequence of the absence of CGM accretion and
of a larger ease for the galaxy degassing, due to the lack of a confining CGM. Figure~\ref{fig5} (right panels) shows $\SigX(R)$ and $\Tx(R)$ of HM$_k^{{\rm noCGM}}$. The
$\SigX(R)$ profile is completely different from that of NGC5129: it is far more peaked in the central galactic region, and too steeply declining outside $R\simeq 10$ kpc
(for plotting purpose $\SigX$ has been rescaled to reach the luminosity of NGC5129,  which requires 
a factor of $\simeq 40$, of course far larger than for the HM models with CGM accretion). 
The temperature is decreased at all radii, and the $\Tx(R)$ profile takes a flatter shape: it
lacks the characteristic bump feature, and a much less pronounced maximum is present  closer to the galactic center (at $R\simeq (1-2)$ kpc).
The lower temperature values, and in particular the steady decline in $\Tx(R)$ outside $R=2-3$ kpc,
are a consequence of the missing confinement effect of the CGM, and the lack of  gravitational compression work done by accretion.
Indeed, there is observational evidence that the shape of $\Tx(R)$ in the outer galactic regions
is sensitive to the presence of an intracluster or intragroup medium (e.g., K20). 
Finally, notwithstanding the decrease in $\Tx$ at all radii, at the center the temperature of HM$_k^{{\rm noCGM}}$ remains slightly larger than observed.

\subsection{Comparison with more galaxies in the $Chandra$ Galaxy Atlas}\label{cga}

The detailed comparison of the previous Section~\ref{profiles} concerned an ETG representative of the HB class of temperature profiles,
the  most commonly observed. We extend here the comparison of the most successful model to reproduce NGC5129 (HM$_k^{{\rm new}}$) to more ETGs of the HB class.
  This analysis will strenghten or weaken the significance of the  results in Section~\ref{profiles}.

\begin{figure}
  \includegraphics[width=1.1\linewidth,  keepaspectratio]{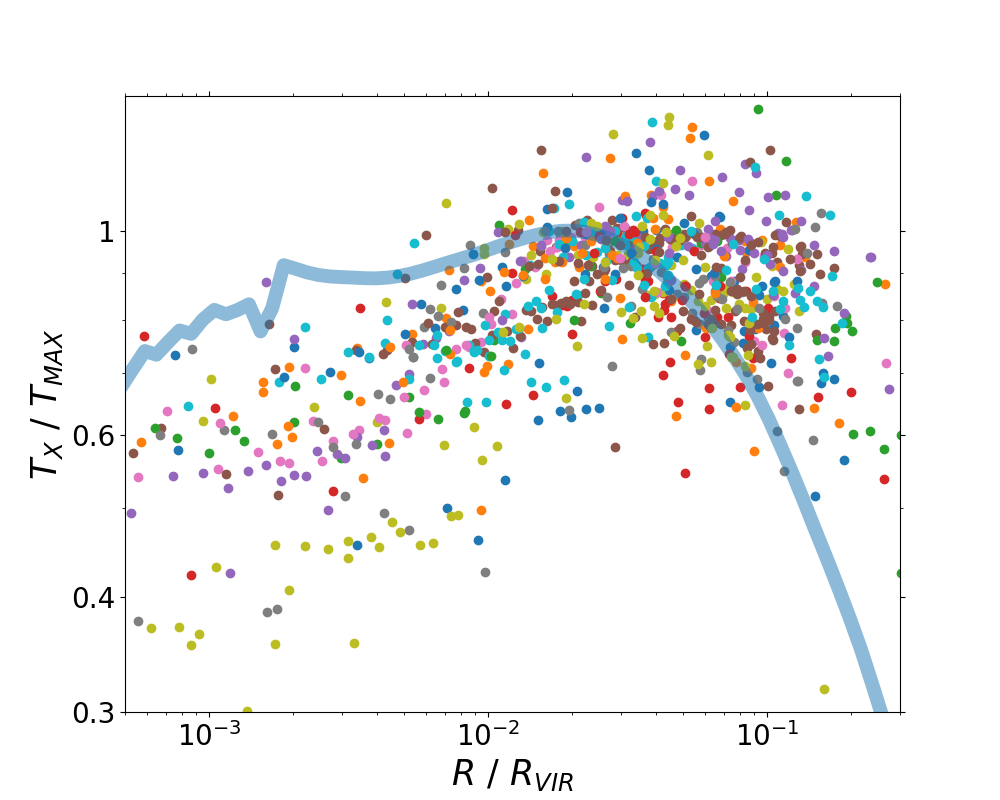}
  \caption{Model HM$_k^{{\rm new}}$ at 7 Gyr (light blue solid line) compared with the temperature data for ETGs
    of the HB class (26 galaxies) and of the double-break class (4 galaxies); each galaxy is plotted with a different colour.
    The temperatures are scaled by $T_{\rm MAX}$, the maximum temperature value, and the galactocentric distances to
    $R_{\rm VIR}$, the virial radius, both determined as detailed in Section~\ref{cga}. This plot is based on Figure 6 in K20, from where the
    observed temperatures are taken.}
\label{fig6}
\end{figure}

Figure~\ref{fig6} shows the temperature profile of HM$_k^{{\rm new}}$ together with all those classified of the HB (26 ETGs) and double-break\footnote{The double-break type shows a $\Tx(R)$ profile that
is falling at small radii, rising at intermediate radii until the peak of the broad bump, and falling again at
large radii. This and the HB types comprise 50 per cent of the K20 sample (30 out of 60 galaxies).} (4 ETGs) types by K20;
the plot is based on Figure 6 in K20.
For a proper comparison, each profile is scaled by its $T_{\rm MAX}$, the maximum temperature of the
best fitting model reproducing it; the radial scale is normalized for each galaxy to its fiducial virial radius $R_{\rm VIR}$, determined from
the average hot gas temperature, as in the relation  from  Helsdon \& Ponman (2003) used by K20.
For HM$_k^{{\rm new}}$ this relation gives $R_{\rm VIR}=0.75$ Mpc, close to $R_{\rm VIR}=0.73$ Mpc of NGC5129 (K20).
The model, plotted with a light blue curve, falls within the distribution of observed points and, even after the scaling that was not applied in Figures~\ref{fig4} and~\ref{fig5},
 shows again a bump located in the radial range where it appears for observed ETGs.
Within $R\approx 0.004 R_{\rm VIR}$, the observed  temperatures show a large scatter, and the model lies on the upper
envelope of the distribution of points. Therefore the result of a temperature larger than observed 
in the central region,  evidenced by the comparison with NGC5129, cannot be discarded as due to some peculiar properties of
this galaxy, but looks like a feature of the numerical model. We note that different estimates for the value of $R_{\rm VIR}$ produce a horizontal shift of the
$\Tx(R)$ profile, and in particular a reduction of $R_{\rm VIR}$ for the model would shift its $\Tx(R)$ to the right, alleviating the discrepancy; however,
a discrepancy was also present in the analysis of  Section~\ref{profiles} (in Figures~\ref{fig4} and~\ref{fig5}) where the radial scale is fixed, and the comparison 
of a model with observations is direct.

\begin{figure}
  \includegraphics[width=1.1\linewidth,  keepaspectratio]{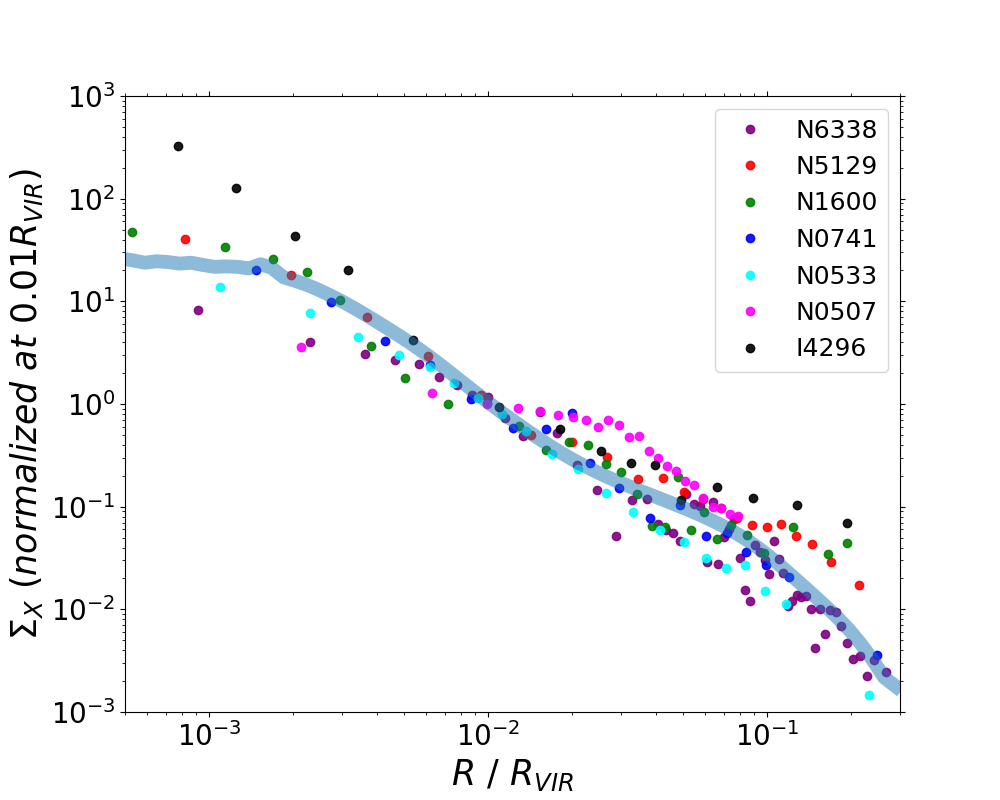}
  \caption{Model HM$_k^{{\rm new}}$ at 7 Gyr  (light blue solid line) compared with the surface brightness profiles for
    7 ETGs with $\Lk$ similar to that of HM models (from K19); the names of these galaxies are reported in the upper right legend. The galactocentric distances are scaled 
by the respective $R_{\rm VIR}$, and each  $\SigX(R)$ is scaled by its value at  $0.01 R_{\rm VIR}$. See Section~\ref{cga} for more details.}
\label{fig7}
\end{figure}

A similar comparison of $\SigX(R)$ of HM$_k^{{\rm new}}$ with that of ETGs in the HB class is more difficult.
In fact, while the temperature profiles could be classified into specific types, the analog classification for $\SigX(R)$ is not available;
indeed, the brightness profiles seem to vary more than the temperature ones. Also, it is more uncertain how to compare them
for a large sample, because the choice of a proper scaling is not straightforward; for example, the brightness peak lies at the center,
and then suffers  from observation-dependent biases (as galaxy distance, exposure, etc.). For these reasons, we first selected those ETGs in K20
with $\Lk$ similar to that of the HM models, i.e., in a range from 4.2 to 7$\times 10^{11}L_{\rm K,\odot}$; of
the ten resulting galaxies, all turned out to belong to the HB temperature class,  except for one that was excluded.
Two ETGs  with a poorly known brightness profile in the  $Chandra$ Galaxy Atlas (K19) were further excluded. The remaining seven galaxies are shown in Figure~\ref{fig7}, together
with model HM$_k^{{\rm new}}$ in light blue; here the galactocentric distances $R$ are measured again in units of $R_{\rm VIR}$, and $\SigX(R)$ of each galaxy is normalized
by its value at the intermediate radius $0.01 R_{\rm VIR}$.  The model compares well with observations over the whole radial range, and represents a reasonable average
for the normalized $\SigX(R)$ profiles. Two galaxies stand out for deviations from the general behavior:
one  (IC4296) shows a steep central increase of $\SigX(R)$  due to an AGN that is 
producing a bright nuclear radio and X-ray source; the other (NGC507) outside of $10^{-2}R_{\rm VIR}$ presents a brightness ``excess''
due to complex substructures in its halo, produced by a radio lobe, sloshing motions and interactions with the nearby galaxy NGC499
(K19, Brienza et al. 2022). 

\vspace{5truemm}

\vspace{5truemm}

\section{Summary, discussion and conclusions}\label{conclu}

In this work we compared the X-ray properties of the gas flow models of C22 with those observed for ETGs and collected in recent,  large and homogeneous studies
based on $Chandra$ data.
The simulations of C22 were conducted with the high-resolution 2D hydrodynamical code MACER (G19a), for a set of realistic galaxies
with three representative stellar masses; for each mass, three galaxy orbital structures were considered:
the non-rotating case, the isotropic rotator, and an intermediate case with a radially declining ordered rotation.  Mass sources are provided by
stellar mass losses, and by a cosmologically motivated time-dependent mass accretion rate imposed at the outer boundary of the numerical grid.
Star formation, that takes place especially in the central gas disk, triggered by the Toomre instability,
acts as a sink and source of mass, the latter due to SNII explosions. Finally, SMBH accretion causes a self-consistently determined AGN feedback, both
radiative and mechanical (due to AGN winds).

First, we compared with observations in the 0.3--8 keV band the global properties of the hot gas: the luminosity  and
the average temperature for the whole galaxy, $\Lx$ and $\avTx$, and those within
$5\Reff$, $\Lxf$ and $\avTf$. The models stay within, and cover most of,  the observed distribution in the diagnostic planes 
$\Lx-\Lk$, $\avTf-\Lk$, $\Lx-\avTx$ and $\Lxf-\avTf$. The observed trend of more massive ETGs hosting on average more luminous and
hotter halos is also reproduced. For each stellar mass, more rotating models
are less X-ray luminous, reinforcing previous results obtained for different galaxy structures, and also in absence of AGN
feedback and CGM accretion. At medium and high galaxy mass (for the MM and HM families) $\avTf$ is lower in rotating models, while the opposite
is shown by the LM family. While $\avTf$ is sensitive to galactic properties, as rotation, and to the presence of circumgalactic accretion, that increases
  its value,   $\avTx$ instead depends mostly on the galaxy mass. Finally, CGM accretion can determine a large difference in $\Lx$ for models of same $\Lk$.
The general agreement  of the C22 models with the observed global properties is not
trivial, given that the input physics was chosen independently of the aim of reproducing the X-ray observations, and ranges from the galaxy dynamical
structure and stellar evolution properties to a self-consistent description of the AGN feedback (within the limits of the central grid resolution of 25 pc),
to cosmological gas accretion.

In a second step  we compared the radial profiles of the surface brightness $\SigX(R)$  and of the projected temperature $\Tx(R)$ with those of
the X-ray bright elliptical NGC5129, that is representative of the most common HB class of temperature profiles (K20). The $\Lk$, $\Reff$, $\sigma_0$ and flattening
of NGC5129 turned out to be similar to those of the HM models.
These present three typical $\SigX$ shapes: the non-rotating HM$_0$ model has the most peaked profile, while ordered rotation creates a
central flattening that becomes more pronounced for larger rotation. The $\SigX$ shape of the HM$_0$  and of the mildly rotating HM$_k$ models
follows quite well that  observed; a close match requires a model $\SigX$ up-scale (of a factor of $\simeq 2.5$) that
could be produced by a uniform increase of the gas density by $\simeq 50\%$, as could be plausibly obtained  
for models tailored on NGC5129. The $\Tx$ profile of the model  looks instead problematic
inside $R\approx 20 $ kpc, where it is larger than in NGC5129 and in some models
shows unobserved spatial fluctuations; outside of 20 kpc, instead, $\Tx$ is smooth and closer to the observed values and shape.
  
In order to better understand the origin of the $\Tx$ profile, additional simulations were performed  for the more satisfactory HM$_k$ model,
keeping all its properties fixed. A model with a larger infall speed for the CGM
(HM$_k^{{\rm new}}$) turned out to compare better than HM$_k$ with the observations of NGC5129: 
at an age close to that of this galaxy, $\Tx(R)$ within $20 $ kpc is smoother and lower, and outer of $30$ kpc it increases slightly, all features that 
bring the model closer to observations. However, $\Tx(R)$ of HM$_k^{{\rm new}}$ remains larger than observed in the central region.
A model without CGM accretion was also studied, finding that its $\Lx$, $\Lxf$, $\avTx$ and $\avTf$ still fall within the observed range, but its $\SigX(R)$ and
$\Tx(R)$ totally fail to reproduce those of NGC5129:  $\SigX(R)$ becomes much more peaked, 
$\Lx$ is far below that of the galaxy; $\Tx(R)$ is  decreased at all radii, especially outside $R\simeq 3$ kpc,
and lacks the characteristic bump.

To extend the validity of the previous results, the more successful model HM$_k^{{\rm new}}$ was compared 
  with other well studied $Chandra$ ETGs of the HB class. What found in  the detailed comparison with NGC5129
  was confirmed: the model represents an average of the observed $\SigX(R)$ profiles normalized by their value at $0.01R_{\rm VIR}$;
it also falls within the observed temperature values, but it  lies on their upper envelope within $\simeq 10^{-2}R_{\rm VIR}$. 
  
In conclusion, the C22 models are generally successful in reproducing the X-ray observations. However, 
in the central region the model temperature appears systematically larger than observed by $\simeq 30-40\%$, 
as evidenced by the specific analysis based on NGC5129 and extended to all the HB galaxies. No simple solution
to this discrepancy has been found, but a few possibilities can be discussed.
First, the presence of temporal and spatial fluctuations in the model $\Tx(R)$ within a few kpc radius, that contrasts with the monotonic, and constant in time, 
$\Tx(R)$ decline at large radii, indicates that the AGN feeback may have an effect too strong within the central region. Therefore, 
one possibility is that the input physics of the AGN feedback should be revised  in the values of some parameters.
However, the mass, momentum and energy injected by the AGN winds cannot be adjusted arbitrarily,
because they obey to physical balance relations that cannot be violated (Ostriker et al. 2010), and that are implemented in the MACER models.
In addition, the AGN feedback efficiencies cannot be reduced much, to prevent an excessive growth of the SMBH.

A second possibility is suggested by the fact that the model $\SigX(R)$ and $\Tx(R)$ are overall close to those observed for NGC5129.
Under the assumption that the simulations are producing a pressure profile consistent with the real one, it is interesting
to check what are the consequences of keeping this same pressure profile but with
the  temperature reduced by the required factor of $\alpha\simeq 1.4-1.6$ within $\simeq 20$ kpc. This would increase the density by the same factor, and 
$\SigX$ by a factor of $\alpha^{3/2}\simeq 1.7-2$, for an emissivity proportional to $\rho^2{\sqrt{T}}$, and by $\alpha^{2}\simeq 2-2.6$,  for
a cooling function approximately constant with the temperature (as reasonable for a temperature within $40\%$ of $kT\simeq 1$ keV).
The resulting  increase of $\SigX$ would be of a factor curiously close to that required in Section~\ref{profiles} to shift the model $\SigX$ upward and reach
that of NGC5129. We recall that in the simulations the gas temperature is a derived quantity, obtained as the ratio between the gas pressure
and density; thus, if for some reason the density in the simulations is lower than in NGC5129, an overestimate of the temperature naturally follows.
Of course, it should be checked whether the same argument can be applied to other ETGs in Figure~\ref{fig6} before this possibility
 can be proposed as a plausible solution.

A third effect that could lower the central $\Tx(R)$ by 30--40\% would be present (but missing in the simulations) if the ISM is multiphase
in a way to require a 3D description. Indeed, colder gas phases have been observed in ETGs, especially if they are central dominant
galaxies in groups and clusters (e.g., Werner et al. 2014, O'Sullivan et al. 2018), and multiphase gas has been found by 3D simulations (e.g., Gaspari et al. 2017, Guo et al. 2023).
In Appendix B.1 we provide simple formulae for the expected change in $\Lx$ and $\Tx$ produced by inhomogeneities  in pressure equilibrium with their surroundings.
Changes in the sought direction and of the required size can be easily originated. For example, from Equation
(B6), a two-phase inhomogeneous gas, with a density larger by a factor $r=5$ in a volume fraction $v=0.1$, has an emission weighted temperature reduced to $\simeq 0.6$
that of its homogeneous counterpart, and a luminosity 1.7 times larger; for $r=3$,  increases in $\Lx$ of 30\%, and decreases in $\Tx$
of 25\% are obtained for a broad range of $v\approx 0.1-0.6$ (these estimates adopt a cooling function roughly
independent of the temperature).  Therefore, 3D density inhomogeneities would reduce $\Tx$ of the
models, and they are especially expected  in the central regions (e.g., Guo et al. 2023), where the model  $\Tx$  appears too large.

Finally, some uncertainties could affect also the measured $\Tx(R)$. For example, from $Chandra$ data of NGC5129, Bharadwaj et al. (2014) find an increasing profile from
$k\Tx=0.9$ keV at a few kpc, to  $k\Tx=1.2$ keV between 30 and 40 kpc; Nugent et al. (2020) also found $k\Tx=0.85$ keV between 2 and 3 kpc radius,
and a peak value of 1.25 keV between 10 and 20 kpc. 
These temperatures are larger than those in Figures~\ref{fig4} and~\ref{fig5}, and closer to the model ones; however, the authors above measured larger temperatures also
outside the peak position: for example, at 60 kpc radius, 
$k\Tx=1$ keV (Bharadwaj et al. 2014), and 1.1 keV (Nugent et al. 2020), while $k\Tx\simeq 0.85$ keV in the model. In addition, as shown in
 Section~\ref{cga},  within $\simeq 10^{-2}R_{\rm VIR}$  the model $\Tx(R)$ seems larger than for most ETGs of the HB class, and thus 
possible measurement uncertainties for NGC5129 cannot be a general solution.

The analysis in this work has shown the potential of a close comparison between the results of high resolution hydrodynamical simulations and
data products from X-ray observatories as $Chandra$. In particular, the C22 MACER exploratory set of models
highlighted the importance of CGM accretion to accomplish an agreement with observed results, thanks to its effect of 
enhancing $\Lx$ (and also $\avTx$, to a lower extent), of producing a radially extended X-ray surface brightness profile and large temperature values
in the outer galactic region. The comparison also highlighted a small but systematic discrepancy in the temperatures inside (1--2)$\Reff$, that
calls for simulations even more closely tailored onto observed galaxies, and/or for a
wider exploration of the parameters describing AGN accretion and feedback, and/or for 3D simulations capable to fully account for multiphase effects.

\bigskip
\bigskip

\begin{acknowledgements}
The referee is thanked for useful suggestions. LC and SP acknowledge support from the Project PRIN MUR 2022 (code 2022ARWP9C) ‘Early Formation and Evolution
of Bulge and HalO (EFEBHO)’, PI: M. Marconi, funded by European Union – Next Generation EU (CUP J53D23001590006). Simulations
analyzed in this work were performed with the Princeton Research Computing resources at Princeton  University, which is a consortium
of groups including the Princeton Institute for Computational Science and Engineering and the Princeton University 
Office of Information Technology’s Research Computing department.
\end{acknowledgements} 

\bigskip
\medskip

\onecolumngrid
\appendix

\onecolumngrid

\section{Radiative transfer and computation of observable quantities in the X-rays}
\label{transf}
 We describe here the procedure adopted to compute the emergent emission and then the surface
brightness $\SigX$ and the projected temperature $T_X$ of the model galaxies.
In the C22 MACER simulations, cold gas, or even a high surface density circumnuclear disk, can be present in the central regions,
and these can be opaque to X-rays. In addition, wherever star formation takes place, it is associated with a metal enrichment that
contributes to modify the X-ray emission and the absorption. X-ray absorption can be significant especially when looking through 
the disk in an edge-on view, as supposed in this work.

\subsection{Transmission of the spectrum}

We calculate first the radiative transfer along the line of sight, by post-processing the hydrodynamical simulation data. The radiative transfer equation
(e.g., Chandrasekhar 1960) reads as:
\begin{equation} 
     {d\Inu\over ds} = - \alphanu\Inu + \jnu,
     \label{eq:radiative-transfer}
\end{equation}
where $\Inu$ is the radiation field specific intensity we aim to evaluate, and $s$ is the distance measured along an arbitrary line of sight starting from some origin.
$\jnu$ and $\alphanu$ are respectively the emission and absorption coefficients per unit volume, and are derived from the atomic processes as described below (Appendix A.2);
for simplicity we ignore scattering processes. The integration of Equation (\ref{eq:radiative-transfer}) leads to the formal solution 
\begin{equation} 
     \Inu(\taunu) =  \Inu(0)\,{\rm e}^{-\taunu} + \int_0^{\taunu}\Snu(\taunup)\, {\rm e}^{-(\taunu - \taunup)}\,d\taunup,
     \label{eq:radiative-transfer-integrated}
\end{equation}
where
\begin{equation} 
\Snu\equiv {\jnu\over\alphanu},\quad \taunu\equiv\int_0^s \alphanu(\spr)\, d\spr
\label{eq:source-function}
\end{equation}
are respectively the source function and the optical depth along the line of sight measured from the origin. Ignoring the background radiation $\Inu(0)$, and recasting the integration in terms of
$s$, Equation (\ref{eq:radiative-transfer-integrated}) gives the specific intensity emerging at $s$ due to emission and absorption processes of the material between $0$ and $s$:
\begin{equation} 
     \Inu (0,s) =  \int_0^ s \jnu(\spr)\,{\rm e}^{-\taunu(\spr, s)}\, d\spr,
     \quad  
     \taunu(\spr,s) \equiv \int_{\spr}^ s \alphanu (s^{\prime\prime})\, d s^{\prime\prime},
     \label{eq:radiative-transfer-final}
\end{equation}
where the meaning of  $\taunu(\spr,s)$ is obvious. 

\subsection{Atomic Processes}

Here we show how the emission and absorption coefficients $\jnu$ and $\alphanu$ were derived, starting from the gas temperature $T$, the 
hydrogen and electron number densities $n_{\rm H}$ and $n_e$, and the chemical composition, that were derived for the gas during the simulations.
We recall that the adopted metal abundances are $Z_*=1.5Z_{\odot}$, where $Z_{\odot}=0.0134$  (Asplund et al. 2009),  for the stellar population
of the massive ETGs, and $Z_{\rm CGM} = 0.15 Z_{\odot}$ for the accreting CGM (C22).
During evolution, the ISM is enriched by the nucleosynthetic yields of SNIa's and SNII's, the latter injected by new star formation and transported
around mostly by AGN winds, and is diluted by CGM accretion (Gan et al. 2019b, Pellegrini et al. 2020).

We first evaluated the frequency-dependent emissivity $\epsilon_\nu$, i.e., the radiated power per unit volume per unit frequency:
\begin{equation}
\epsilon_\nu =  n_{\rm H} n_e \Lambda_\nu,
\label{eq:emissivity}
\end{equation}
where $\Lambda_\nu$ is the cooling function per unit frequency, and depends on the temperature $T$ and on the chemical composition.
$\Lambda_\nu$ was evaluated by using the software package \texttt{ATOMDB} (version 3.0.9;  assuming collisional ionization equilibrium).
Assuming that the X-ray emission is isotropic, the emission coefficient $\jnu$ is simply given by 
\begin{equation} 
\jnu =  {\epsilon_\nu\over 4\pi}.
\label{eq:emission-coefficient}
\end{equation}

The absorption coefficient $\alphanu$ is a function of gas density and metal abundance. For the solar abundance,
we adopt the photoelectric absorption cross-section $\sigma_{\rm \nu,solar}=\sigma_{\rm \nu, H+He}+\sigma_{\rm \nu, metal}$ of Morrison \& McCammon (1983),
in which H and He contribute $\sigma_{\rm \nu, H+He}$ and dominate the cross section for soft X-ray photons ($E\lesssim0.5$ keV), while the absorption of hard X-ray photons is mainly
contributed by metals (elements heavier than He).  Assuming that the X-ray absorption by the metal-rich ISM can be written as a linear function of metallicity $Z$ (in units of solar metalliticy
$Z_{\odot}$), we adopt:
\begin{equation} 
\sigma_{\rm \nu}=\sigma_{\rm \nu, H+He}+ Z\times\sigma_{\rm \nu, metal}.
\label{eq:absorption-cross-section}
\end{equation}
The absorption coefficient is then given by
\begin{equation}
\alphanu =  n_{\rm H} \sigma_\nu.
 \label{eq:absorption-coefficient}
\end{equation}

\subsection{X-ray surface brightness and luminosity, temperature profile and average temperatures, for the models}

Using Equations (\ref{eq:radiative-transfer-final}), (\ref{eq:emission-coefficient}), and (\ref{eq:absorption-coefficient}), we can derive the observed properties of the models in the X-ray band,
that in this work we take as the broad $Chandra$ band of 0.3--8 keV. To evaluate 
the radiative transfer, we perform the calculations in a fashion of ray-tracing, i.e., we integrate Equation (\ref{eq:radiative-transfer-final}) along the direction from the emitter to the observer.
The radiation in and behind the circumnuclear gaseous disk is attenuated whenever the disk is optically thick; this affects especially the soft X rays. In the MACER simulations, spherical coordinates
(with a logarithmic radial grid to cover a large dynamical range, from $r=2.5$ pc to $\sim 250$ kpc in the highest resolution runs) make the integration along the straight line $s$ not immediate.
We then interpolated the gridded data from the original spherical coordinates onto a new cylindrical coordinate system $(z, \tilde r, \varphi)$,
with the $z$-axis along the line of sight pointing toward the observer, so that the numerical integration reduces to a summation along the $z$-axis of the new cylindrical coordinate system.
Moreover, to preserve the high resolution in the galactic central regions, we adopted logarithmic spacing in both the $z$ and $\tilde r$ directions. We thus integrated Equation
(\ref{eq:radiative-transfer-final}) for fixed $(\tilde{r}, \varphi)$ in the projection plane, with $s$ replaced by $z$ and spanning the whole interval $(\zmin,\zmax)$ covered by the numerical
grid, to obtain the monocromatic 2D brightness distribution $\Signu$, and then the X-ray surface brightness $\SigX$ after integration over the energy band of interest:
\begin{equation} 
\Signu(\tilde{r},\varphi)=4\pi\Inu(\zmin,\zmax),\quad  
\SigX(\tilde{r},\varphi) = \int_{\numin}^{\numax}\Signu(\tilde{r},\varphi)\,d\nu.
     \label{eq:xray-surface-brightness}
\end{equation}
%
The total X-ray luminosity $\Lx$ is evaluated by integrating $\SigX$ over the surface area $dA=\tilde{r}\,d \tilde{r} d\varphi$ on the sky plane, i.e.
\begin{equation}
          \Lx =  \int\SigX(\tilde{r},\varphi)\, dA.
      \label{eq:xray-luminosity}
\end{equation}
If the integration is performed within a radius of $5\Reff$, one obtains $\Lxf$, the luminosity within a cylinder with axis along the line of sight and with basis a circle of radius $5\Reff$.

In the analysis of X-ray observations, the temperature $T_X$ is derived from the X-ray spectra, and it is an emission weighted quantity, with good approximation
(e.g., K19, Truong et al. (2020). Therefore, as a proxy for $T_X$, we evaluate the projection of
the gas temperature $T$ along the line of sight, weighting it by the X-ray emission after absorption:
\begin{equation} 
     \Tx (\tilde{r},\varphi)=
     {\int_{\numin}^{\numax}\,d\nu\int_{\zmin}^{\zmax} T(z)\,\jnu(z)\,{\rm e}^{-\taunu(z,\zmax)}\,dz\over
       \int_{\numin}^{\numax}\,d\nu\int_{\zmin}^{\zmax}\jnu(z)\,{\rm e}^{-\taunu(z,\zmax)}\, dz}. 
      \label{eq:apparent-temperature}
\end{equation}
Finally, we calculate the circularized surface-brightness profile $\SigX(R)$ and the circularized 
surface-brightness-weighted temperature profile $\Tx(R)$ as angle averaged quantities over the annulus $R-\Delta R/2 < \tilde{r} < R+\Delta R/2$:
\begin{equation} 
 \SigX (R) =  
     \displaystyle{
     \int_0^{2\pi}\int_{R-\Delta R/2}^{R+\Delta R/2}\SigX(\tilde{r},\varphi)\,dA
     \over
     2\pi R\Delta R
     },
     \quad
     \Tx (R) =  
     \displaystyle{
     \int_0^{2\pi}\int_{R-\Delta R/2}^{R+\Delta R/2}\Tx (\tilde{r},\varphi)\,\SigX(\tilde{r},\varphi)\,dA
     \over
     \int_0^{2\pi}\int_{R-\Delta R/2}^{R+\Delta R/2} \SigX(\tilde{r},\varphi)\,dA
     }.
     \label{eq:projected-temperature-TR}
   \end{equation}
   For the comparison with the average temperature observed within a given aperture (e.g., 5$\Reff$), we define for the models:
   \begin{equation}
        \avTx(R)= {\int_0^R \SigX(R') \, \Tx(R') \, R'\, dR' \over \int_0^R \SigX(R') \, R'\, dR'}     .
     \label{eq:aver-TR}
   \end{equation}
   The average temperature over the whole galaxy is indicated with $\avTx$, and that within $5\Reff$ (i.e., $\avTx (5\Reff)$) with
   $\langle T_{\rm X,5} \rangle$.

   \section{Effects of a Multiphase Medium on observed temperature and luminosity}

   As discussed in Section 5, the emission weighted temperature can be reduced in a multiphase medium, with a reduction factor comparable
   to that needed to bring the model profiles in a better agreement with the observed ones. The C22 simulations are 2D, and it is reasonable to
   expected that in a high-resolution 3D simulation the number of inhomogeneites for unit volume would increase, especially in the inner galactic
   regions, where AGN feedback heats and compresses the gas (discussed in B.1). Another possible source of ISM inhomogeneites resides in
   the process of 3D fragmention of the cold gaseous disk (discussed in B.2).

  \subsection{Emission-weighted average temperature and luminosity of a multiphase medium}
 
  We present here a simple model to estimate the effects of a multiphase medium on the luminosity and on the luminosity-weighted
  temperature measured when averaging over a sufficiently large volume. We consider first a volume $V$, filled with a gas of uniform
  density $\rho$, total mass $M=\rho V$, and uniform temperature $T$; the pressure is $p=k_B\rho T/(\mu m_p)$, and the total internal energy is
  $U=(3/2)pV$. The emission per unit volume is assumed to be $A\rho^2 T^{\lambda}$, with $A$ and $\lambda$ given
  constants; in this way, the total luminosity of the gas is $L= A\rho^2 T^{\lambda} V$. 
  We now consider what luminosity $\avL$ and emission weighted temperature $\avT$ are
  obtained if the same volume $V$ contains the same amount of gas $M$, but the gas is distributed in
  $N$ different phases of density $\rhoi$, each occupying the volume $\Vi$; moreover, we assume pressure equilibrium between the different phases,
  i.e., $p_1=p_2=...=p_N$, and we also require that the total internal energy of the non-homogeneous configuration is the same as for the
  homogeneous system. We define the normalized quantities
  \begin{equation}
\Vit\equiv{\Vi\over V}, \quad\rhoit\equiv{\rhoi\over\rho}, 
\quad 
\end{equation}
and we have that:
\begin{equation}
  \sum_{i=1}^N\Vit=1, \quad \sum_{i=1}^N\rhoit\Vit=1, \quad \Ti= {T\over\rhoit}.
\label{eq:multi1}
\end{equation}
Therefore, for the multiphase gas, the luminosity $\avL $ and the temperature $\avT $ are:
\begin{equation}
\avL \equiv \sum_{i=1}^N L_i = L\times\sum_{i=1}^N \rhoit^{2-\lambda}\Vit,\quad 
\avT \equiv { \sum_{i=1}^N L_i T_i \over \avL} =T\times{\sum_{i=1}^N \rhoit^{1-\lambda}\Vit\over\sum_{i=1}^N \rhoit^{2-\lambda}\Vit}; 
\label{eq:multi2}
\end{equation}
as expected, for $\lambda=1$, $\avL=L$ and $\avT = T$.
In order to obtain a numerical estimate, we consider the simple case of a two-phase medium, where we introduce the density ratio between the high and the
low density phases, and the corresponding volume ratio:
\begin{equation}
r\equiv{\rhohi\over\rholo}, \quad v\equiv{\Vh\over\Vl}.
\label{eq:multi3}
\end{equation}
From the general relations it follows that 
\begin{equation}
  \tilde\rho_{\rm high}=r \tilde\rho_{\rm low},  \quad
  \tilde V_{\rm high}=v \tilde V_{\rm low},\quad
\tilde\rho_{\rm low}={1+v\over 1+rv},\quad   
  \tilde V_{\rm low}={1\over 1+v},
\label{eq:multi4}
\end{equation}
so that 
\begin{equation}
\avL=L\times {(1+v)^{1-\lambda}(1+r^{2-\lambda}v)\over (1+r v)^{2-\lambda}},\quad 
\avT=T\times {(1+r v)(1+r^{1-\lambda}v)\over (1+v)(1+r^{2-\lambda}v)}. 
\label{eq:multi5}
\end{equation}
These formulae are used in Section~\ref{conclu} to assess the effect of density inhomogeneities on the luminosity and temperature measured for a region
that hosts them. For $\lambda<1$, at any $v$ the luminosity is enhanced ($\avL>L$), and the temperature is reduced ($\avT<T$),
and these variations  are larger for larger $r$. Note that a density ratio of $r$ corresponds to a temperature ratio of $1/r$; thus, if $r$ is small (say $\la 5$),
and the temperature of the homogeneous configuration is such that the gas emits  in the X-ray band,  the
temperatures of the different  phases remain within a range where most of the emission is in the X-ray band, and Equation (B6) can be used to
obtain the multiphase $\Lx$ and $\Tx$.

\subsection{Fragmentation in a central gaseous disk}

We give here a simple argument by which 3D instabilities are to be expected  in the cold rotating disk subject to Toomre instability. As implemented in the
code, at radius $R$ the disk becomes locally unstable when 
\begin{equation}
\QT(R)={\cD (R)\kr(R)\over \pi G\Sigma (R)}<1, 
\end{equation}  
where $\Sigma (R)$, $\cD(R)$, and $\kr(R)$ are respectively the local gas surface density, speed of sound, and radial epicyclic frequency. Assuming for simplicity a
roughly constant disk rotational velocity $\Vrot$,
then $\kr=\sqrt{2}\Vrot/R$, and so for a marginally stable disk at $R$
\begin{equation}
\SigmaT(R)\equiv{\cD (R)\kr(R)\over\pi G} = {\sqrt{2}\over\pi}{\cD(R)\Vrot\over G R}. 
\end{equation}  
We now assume that a region of radius $\rJ$ around $R$ collapses due to Jeans instability, so that from the identities
\begin{equation}
\MJ\equiv {4\pi\over 3}\rJ^3\rhoJ=\pi\rJ^2\SigmaT(R),\quad \rJ=\sqrt{\pi\over 4G\rhoJ}\cD(R),
\label{eq:massj}
\end{equation}
we can express the various quantities in terms of the properties of the disk as
\begin{equation}
\rJ={\pi\cD(R)^2\over 3 G\SigmaT(R)} ={\pi^2\cD(R) R\over 3 \sqrt{2}\Vrot},\quad
\rhoJ={9 G \SigmaT(R)^2\over 4\pi\cD(R)^2}={9\Vrot^2\over 2\pi^3 G R^2},\quad
\MJ={\pi^3\cD(R)^4\over 9G^2\SigmaT(R)}={\pi^4\cD(R)^3 R\over 9\sqrt{2}G\Vrot}.
\label{eq:soljeans}
\end{equation}
Therefore, as an order-of-magnitude estimate, the first of the equations above indicates that a Toomre unstable ring, in a disk with $\Vrot=100$ km s$^{-1}$ and $T_{disk} = 100$ K, would fragment in a number 
$\pi R/\rJ\approx 100$ blobs, that would cool, collapse, and increase their density.

\twocolumngrid

$\,$
\bigskip


\end{document}